\documentclass[aps,prl,twocolumn,amsmath,amssymb,nofootinbib,superscriptaddress,12pt,reprint]{revtex4-1}
\usepackage{times}
\usepackage[pdftex]{graphicx} 
\usepackage{caption} 
\usepackage{subcaption}
\usepackage{dcolumn}
\usepackage{bm}
\usepackage{amsmath}
\usepackage{indentfirst}
\usepackage{float}
\usepackage[colorlinks]{hyperref}
\usepackage{cleveref}
\usepackage[dvipsnames]{xcolor}
\usepackage{qcircuit}
\usepackage{CJKutf8}
\usepackage[ruled]{algorithm} 
\usepackage{algpseudocode}
\usepackage{multirow}

\begin{document}

\title{A hybrid quantum walk model unifying discrete and continuous quantum walks}

\author{Tianen Chen}
\affiliation{Institute of Mathematics, Academy of Mathematics and Systems Science, Chinese Academy of Sciences, Beijing
100190, China}
\affiliation{School of Mathematical Sciences, University of Chinese Academy of Sciences, Beijing 100049, China}
\author{Yun Shang}
\email{shangyun@amss.ac.cn}
\affiliation{Institute of Mathematics, Academy of Mathematics and Systems Science, Chinese Academy of Sciences, Beijing
100190, China}
\affiliation{State Key Laboratory of Mathematical Sciences, Academy of Mathematics and Systems Science, Chinese Academy of Sciences, Beijing 100190, China}

\begin{abstract}
Quantum walks, both discrete and continuous, serve as fundamental tools in quantum information processing with diverse applications. This work introduces a hybrid quantum walk model that integrates the coin mechanism of discrete walks with the Hamiltonian-driven time evolution of continuous walks. Through systematic analysis of probability distributions, standard deviations, and entanglement entropy on fundamental graph structures (2-vertex circles, stars, and lines), we reveal distinctive dynamical characteristics that differentiate our model from conventional quantum walk paradigms. The proposed framework demonstrates unifying capabilities by naturally encompassing existing quantum walk models as special cases. Two significant applications emerge from this hybrid architecture: (1) We develop a novel protocol for perfect state transfer(PST) in general connected graphs, overcoming the limitations of previous graph-specific approaches. A PST on a tree graph has been implemented on a quantum superconducting processor. (2) We devise a quantum algorithm for multiplying $K$ adjacency matrices of $n$-vertex regular graphs with time complexity $O(n^2d_1\cdots d_K)$, outperforming classical matrix multiplication $(O(n^{2.371552}))$ when vertex degrees $d_i$ are bounded. The algorithm's efficacy for triangle counting is experimentally validated through the quantum simulation on PennyLane. These results establish the hybrid quantum walk as a versatile framework bridging discrete and continuous paradigms while enabling practical quantum advantage in graph computation tasks.
\end{abstract}


\maketitle

\vspace{8mm}

\section{\MakeUppercase\expandafter{\romannumeral1}. Introduction}

Quantum walks, as a universal quantum computation model\cite{3,12,13,41,42}, have demonstrated remarkable computational advantages. They achieve exponential speedups in tasks such as the glued trees problem \cite{2} and offer significant improvements in search algorithms \cite{27,29,30,31,32}, combinatorial optimization problems \cite{28}, Hamiltonian simulation \cite{11}, graph isomorphism \cite{33,34,35,36}, the triangle finding \cite{37}, the element distinction \cite{38}, the subset finding \cite{39}, constraint satisfaction problems \cite{40} and so on. 

Quantum walks are primarily classified into discrete and continuous models, corresponding to their classical counterparts \cite{23}. In the realm of classical and quantum control theory, hybrid systems that integrate both continuous dynamic evolution and discrete controls have garnered increasing attention in academia and industry \cite{62,63,64,65,66,67,68}, particularly for fault-tolerant quantum computing \cite{69,70}. Combining discrete quantum walks based on coin controls  \cite{8,9,14,15} with the continuous quantum walks \cite{2} into hybrid quantum walks may bring new discoveries to quantum algorithms and quantum control theory. However, due to the operational differences between coin-driven models and Hamiltonian-based propagation, creating hybrid frameworks remains challenging. 

Underwood et al. introduced a discontinuous quantum walk model that integrates discrete and continuous dynamics via edge coloring of graphs and color-switching mechanism \cite{3}. In this framework, a chromatic subgraph of a specific color is initially selected,  followed by the execution of continuous quantum walk evolution on the retained subgraph. Although this model achieves universal quantum computation through engineered perfect state transfers in continuous quantum walks, its discrete dynamics remain limited to graph-switching mechanisms, omitting the essential incorporation of coin operators, which are both a defining characteristic and functional cornerstone of discrete quantum walks\cite{14,18,19,21,22}. The absence of coin-based state control intrinsically restricts the model's ability to emulate authentic discrete walk features, effectively reducing it to concatenated continuous walks interspersed with chromatic switching. Recent phase-space quantum walk implementations offer alternative hybrid strategies by encoding continuous variables through geometric phase manipulations on circles, lines, or Poincaré disks \cite{1,4}, suggesting new paradigms for synthesizing discrete and continuous walk properties while preserving their intrinsic operational features. However, existing hybrid models either neglect critical discrete elements such as coin operators or constrain continuous evolution to specific geometries, failing to generalize across arbitrary graphs.

The challenge to simultaneously maintaining discrete control (via coins) and continuous dynamics on general graphs motivates our work. We propose a hybrid quantum walk framework that unifies discrete and continuous dynamics by integrating discrete coin operators for controlling coin states while introducing Hamiltonian dynamics on arbitrary graphs. This approach embeds coin operations into continuous quantum walks, enabling simultaneous discrete coin controls and continuous propagation via graph Hamiltonians. Systematic analysis on 2-vertex circles, star graphs, and lines reveals unique dynamical signatures in probability distributions and entanglement entropy, distinguishing our model from conventional quantum walks. This hybrid architecture, combining discrete and continuous quantum walks, achieves highly efficient state transfer and superior computational performance on general graphs. Through the joint engineering of coin operators and continuous quantum walks, it enables perfect state transfer on arbitrary connected graphs, overcoming previous limitations imposed by specific topological structures \cite{21}. For $K$ regular graphs with $n$ vertices and bounded degrees $\{d_1, \cdots, d_K\}$, the architecture implements a quantum adjacency matrix multiplication algorithm with complexity $O(n^2 d_1\cdots d_K)$, demonstrating a clear advantage over the classical state-of-the-art method $(O(n^{2.371552}))$ \cite{49}. Experimental verification on the PennyLane platform confirms its practical effectiveness in solving triangle counting problems \cite{51}, a fundamental task in graph analysis. By establishing coordinated control between discrete coin operators and continuous-time evolution, our hybrid framework successfully integrates discrete and continuous quantum walk paradigms.

This paper is structured as follows. The background information regarding discrete coin-based quantum walks and continuous quantum walks is given in section II. In section III, we introduce the construction method of the hybrid quantum walk model and describe the dynamic evolutions and properties for hybrid quantum walks on 2-vertex circles, star graphs, and lines. In section IV, we explain how to implement the discontinuous quantum walk using our hybrid quantum walk. In section V, we introduce how to achieve perfect state transfer on general graphs using the hybrid quantum walks. In section VI, we present a new quantum matrix multiplication algorithm for regular graph adjacency matrices using the hybrid quantum walk model and apply the algorithm to the triangle counting problem of regular graphs with experimental validation on PennyLane. Finally, a discussion is given in section VII.

\section{\MakeUppercase\expandafter{\romannumeral2}. Discrete and continuous quantum walks}.\label{s2}

Discrete quantum walk models mainly include coin-based quantum walks \cite{14,15,8,9}, Szegedy's quantum walks \cite{16}, and staggered quantum walks \cite{17}. Among these, coin-based quantum walks are a pivotal model in discrete quantum walks. They draw inspiration from  classical random walks, specifically the process of determining movement direction based on coin flipping outcomes. Coin-based quantum walks invlove two Hilbert spaces: the coin space $H_c$ and the position space $H_p$, along with two unitary operators: the coin operator $C$ acting on $H_c$ and the conditional shift operator $S$ acting on $H_c\otimes H_p$. Each step of the quantum walk is characterized by the unitary operator $S(C\otimes I)$, where $C$ depicts the coin flipping, and $S$ illustrates how the outcome of the coin flipping influences the direction of the walk. In a graph $G=(V, E)$, the computational bases for $H_c$ and $H_p$ correspond to the edge set and the vertex set, respectively, that is, $H_c=\mathrm{span}\{|a\rangle: a \in \Gamma\}$, $H_p=\mathrm{span}\{|n\rangle: n \in V\}$, where $\Gamma$ denotes the set of labels for edges $E$. For a given graph, while the conditional shift operator $S$ is typically determined, the coin operator $C$ exhibits significant flexibility, which is a distinctive feature of coin-based quantum walks. Beyond the single-coin model, coin-based quantum walks also include multi-coin models \cite{18} and entangled coin models \cite{19}. Multi-coin models have proven instrumental in various quantum information technologies, such as quantum teleportation \cite{20}, perfect state transfer \cite{21}, and entanglement preparation and distribution \cite{22} underscoring the critical role of the coin operator in coin-based quantum walks..

Similar to discrete quantum walks, continuous quantum walks \cite{2} are quantized versions of classical random walk models—specifically, continuous-time Markov chains \cite{24}. Unlike discrete random walks, continuous-time Markov chains do not involve auxiliary coins but instead use time-dependent probability transition matrices to characterize the walk mechanism. According to Kolmogorov's backward equation \cite{25}, the time evolution of this probability transition matrix is governed by a generator matrix \cite{26}. For an undirected graph with $N$-vertex $G$, the generator matrix is given by $H=\gamma(A-D)$, where $\gamma$ denotes the hopping rate, $A$ is the adjacency matrix of the graph $G$, and $D=\mathrm{diag}(d_1, d_2,\cdots,d_N)$ is the degree matrix. Since the generator matrix $H$ is Hermitian, in continuous quantum walks, $H$ can be used directly as the Hamiltonian to describe the evolution of quantum states. Therefore, the continuous quantum walk operator can be expressed as $U(t)=e^{-iHt}$ acting on the Hilbert space $\mathrm{span}\{|n\rangle: n \in V\}$. Depending on the problem at hand, one may also choose $H=A$ or $H=A-D$. In this article, we adopt the adjacency matrix $A$ as the Hamiltonian to study continuous quantum walks.

\section{\MakeUppercase\expandafter{\romannumeral3}. Hybrid quantum walk model}\label{s3}

\begin{figure}[htbp]
  \centering
  \includegraphics[width=0.45\textwidth]{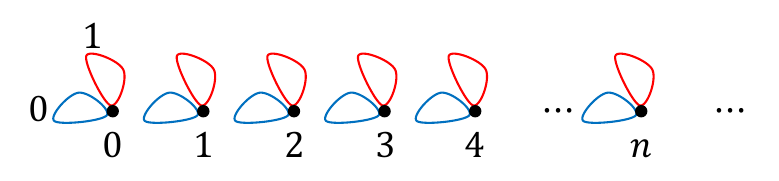}\\
  \caption{The graph $G_0$. The blue and red self-loops with different weights are labeled as $0$ and $1$, respectively. }
  \label{fig:1}
\end{figure}

To construct the hybrid quantum walk model, we begin with the interaction term $H_{JC}=\frac{\hbar g}{2}\sigma_z\otimes\hat{a}^{\dag}\hat{a}$ from the Jaynes-Cummings model Hamiltonian in the large detuning limit\cite{43}. Here, $\sigma_z=|0\rangle\langle0|-|1\rangle\langle1|$ acts on the spin-$1/2$ space, $\hat{a}^{\dag}\hat{a}$ is the photon-number operator for the light mode, and $g$ quantifies the spin-light coupling strength. The associated time-evolution operator $S_{JC}(t)=e^{-iH_{JC}t}$ takes the form
\begin{align}
S_{JC}(t)=\exp\{-i(|0\rangle\langle0|\otimes(\frac{g}{2}\hat{a}^{\dag}\hat{a})-|1\rangle\langle1|\otimes(\frac{g}{2}\hat{a}^{\dag}\hat{a}))t\}.    
\end{align}
This operator acts on the composite Hilbert space \begin{align}
H_c\otimes H_p =\mathrm{span}\{|0\rangle, |1\rangle\}\otimes \mathrm{span}\{|n\rangle: n=0, 1, \cdots \}.    
\end{align}
and can be reinterpreted as a conditional displacement operator for a quantum walk on the Fock-state lattice. Specifically, using the eigenvalue relation $\hat{a}^{\dag}\hat{a}|n\rangle=n|n\rangle$, we rewrite $S_{JC}(t)$ as \begin{align}
S_{JC}(t)=\exp\{-i\sum_{n\in\mathbb{N}}\frac{gn}{2}(|0\rangle\langle0|-|1\rangle\langle1|)\otimes|n\rangle\langle n|t\}.  
\end{align}
which induces phase shifts conditioned on both the spin state $|0\rangle$ or $|1\rangle$ and the photon number 
$n$. To geometrize this dynamics, we associate it with a graph
 $G_0=(V_0, E_0)$ as shown in FIG.~\ref{fig:1}, where $V_0=\{0, 1, 2, \cdots\}$ represents Fock state and $E_0$ contains self-loops at each vertex labeled by $\Gamma_0=\{0,1\}$. 
Every vertex carries two self-loops with weights $\frac{gn}{2}$ (label $0$) and $-\frac{gn}{2}$ (label $1$) corresponding to the diagonal entries of $H_{JC}$. Crucially, $H_{JC}$ is Hermitian but non-unitary, analogous to a graph Laplacian rather than a conventional quantum walk shift operator.

Since FIG.~\ref{fig:1} has only self-loops and no edges between vertices, it can only lead to phase shifts on the amplitudes of the different vertices, and cannot well represent the position changes of walkers caused by the quantum walk. So we need to generalize the above special case to general graphs. To generalize this structure to arbitrary graphs $G=(V,E)$, we decompose $E$ into labeled subsets $\{E_j\}_{j\in \Gamma}$ (not necessarily disjoint) and define the generalized Hamiltonian
\begin{align}
    H=\sum_{j\in\Gamma}|j\rangle\langle j|\otimes S_j 
\end{align} 
where $S_j$ is the (weighted) adjacency matrix for subgraph for $G_j=(V, E_j)$. Hermiticity of $H$ requires each $S_j$ to be Hermitian. The composite Hilbert space becomes 
\begin{align}
H_c\otimes H_p= \mathrm{span}\{|j\rangle: j\in \Gamma\}\otimes \mathrm{span}\{|v\rangle: v\in V\},     
\end{align} with $H$ acting as 
\begin{align}
    H|j, u\rangle=\sum_{u,v\in E_j}w(u, v)|j, v\rangle,
\end{align} where $w(u, v)$ denotes the edge weight. When ${E_j}$ partitions $E$, then $\sum_{j\in\Gamma}S_j$ recovers the  adjacency matrix $A$ of $G$. 

The hybrid quantum walk dynamics on $G$ is then governed by:
\begin{align}
    W(t)=S(t)(C\otimes I)=e^{-iHt}(C\otimes I),
\end{align} 
where $C$ is a unitary operator on $H_c$ and $I$ is the identity operator on $H_p$. 
This formulation generalizes the $JC$-inspired phase dynamics to arbitrary graph structures without restricting
$G$'s connectivity. 

To demonstrate the generality of our framework, we first consider a minimal case where $\Gamma=\{a\}$ (single edge label) and $C=I$. 
The walk operator simplifies to
\begin{align}
W(t)=e^{-i|a\rangle\langle a|\otimes At}=I\otimes e^{-iAt},   
\end{align}
which reduces to a continuous-time quantum walk on graph $G$ when restricted to the position space $H_p$. For the conventional discrete quantum walk with a coin $W=S(C\otimes I)$, where the conditional shift operator $S=\sum_{j\in\Gamma}|j\rangle\langle j|\otimes \widetilde S_j$ and each $\widetilde S_j$ are unitary operators, so there is a Hamiltonian $H=\sum_{j\in\Gamma}|j\rangle\langle j|\otimes S_j$ such that $S=\sum_{j\in\Gamma}|j\rangle\langle j|\otimes e^{-iS_j}=e^{-iH}$. Thus the quantum walk $W=S(C\otimes I)=e^{-iH}(C\otimes I)$ can be regarded as a special case of our hybrid quantum walks. Therefore, both conventional continuous quantum walks and discrete quantum walks with coins can be regarded as special cases of hybrid quantum walks.

\subsection{SU(1,1) Realizations}

\begin{figure}[htbp]
  \centering
  \includegraphics[width=0.45\textwidth]{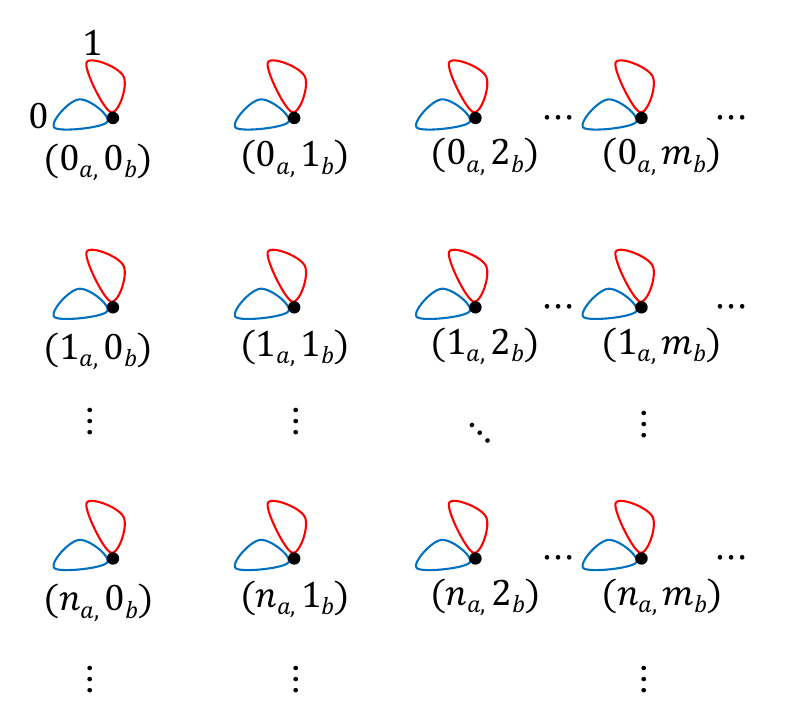}\\
  \caption{The graph $G_0'$. The blue and red self-loops with different weights are labeled as $0$ and $1$, respectively. }
  \label{fig:16}
\end{figure}

A more instructive example emerges from the $SU(1,1)$ Lie algebra realization \cite{1}.
  For the $single-mode$ realization, the generators are $\hat{K}_0=\frac{1}{2}(\hat{a}^{\dag}\hat{a}+\frac{1}{2})$, $\hat{K}_+=\frac{1}{2}(\hat{a}^{\dag})^2$ and $\hat{K}_-=\frac{1}{2}(\hat{a})^2$. Then the conditional shift operator becomes 
\begin{align}
S(t)&=e^{-i\sigma_z\otimes\hat{K}_0t}\nonumber\\
&=\exp\{-i\sum_{n\in\mathbb{N}}\frac{2n+1}{4}((|0\rangle\langle0|-|1\rangle\langle1|)\otimes|n\rangle\langle n|)t\}, 
\end{align}
implementing a hybrid quantum walk  on the graph $G_0$ with vertex-dependent self-loop weights: $w^{\{0\}}_{n}=\frac{2n+1}{4},w^{\{1\}}_{n}= -\frac{2n+1}{4}$, respectively. 
For the $two-mode$ with mode $a$ and $b$, the generators generalize to $\hat{K}_0=\frac{1}{2}(\hat{a}^{\dag}\hat{a}+\hat{b}^{\dag}\hat{b}+1)$, $\hat{K}_+=\hat{a}^{\dag}\hat{b}^{\dag}$ and $\hat{K}_-=\hat{a}\hat{b}$. The shift operator now acts on the extended space $H_{c}\otimes\{H_a\otimes H_b\}$ as
\begin{align}
S(t)&=e^{-i\sigma_z\otimes\hat{K}_0t}\nonumber\\
&=\exp\{-i\sum_{n\in\mathbb{N}}\frac{n+m+1}{2}\sigma_z\otimes|n,m\rangle_{a,b}\langle n,m|t\}. 
\end{align}
which corresponds to a hybrid walk on the bipartite graph
$G_0'$ as shown in FIG.~\ref{fig:16}, each vertex $(n_a, m_b)$ carries two self-loops labeled with 0 and 1 with weights $\frac{n+m+1}{2}, -\frac{n+m+1}{2}$, respectively. 

Next, we will characterize the hybrid quantum walk models on 2-vertex circle, star graphs, and lines.

\subsection{2-vertex circle}

\begin{table*}[htbp]
    \centering
    \begin{tabular}{c|c|c}
    \hline
    \hline
    \textbf{Quantum Walk(QW)}  & \textbf{Parameter Selection} & \textbf{Core Characteristic}  \\
        \hline
    Symmetric Hybrid QW & $a=b=2\omega$ (FIG.~\ref{fig:2}) & Single frequency oscillation without beat frequency effect  \\
         \hline
    Weak Asymmetric Hybrid QW & $a=2\omega$ and $b=2\omega+1$ (FIG.~\ref{fig:3}) & Low frequency envelope modulation, periodic stable bands  \\
    \hline
    Strong Asymmetry Hybrid QW & $a=4\omega$ and $b=4\omega+3$ (FIG.~\ref{fig:4}) & High frequency resonance, dense periodic stable bands  \\
    \hline
    Continuous QW & $\omega$ (FIG.~\ref{fig:5}) & Single frequency oscillation without beat frequency effect  \\
    \hline
    \end{tabular}
    \centering
    \caption{The selection of types and parameters of quantum walks on a 2-vertex cycle, and their dynamic core characteristics.}
    \centering
    \label{tab:1}
\end{table*}

Consider a graph $G=(V, E)$ with $V=\{0, 1\}$ and edge set $E$ forming a bidirectional cycle. 
 Let $\Gamma=\{0, 1\}$ label edges with weights $a$ and $b$. The walk space becomes $H_c\otimes H_p=\mathrm{span}\{|0\rangle, |1\rangle\}\otimes \mathrm{span}\{|0\rangle, 1\rangle\}$, with the evolution Hamiltonian $H=|0\rangle\langle0|\otimes aX+|1\rangle\langle1|\otimes bX$, where $X$ is the Pauli $X$ matrix. The time-evolution operator decomposes as 
\begin{align}
S(t)=&|0\rangle\langle0|\otimes(\cos{(at)}I-i\sin{(at)}X)\nonumber\\
+&|1\rangle\langle1|\otimes(\cos{(bt)}I-i\sin{(bt)}X),    
\end{align}
yielding the walk operator
\begin{align}
W(t)=&(|0\rangle\langle0|\otimes(\cos{(at)}I-i\sin{(at)}X)\nonumber\\
+&|1\rangle\langle1|\otimes(\cos{(bt)}I-i\sin{(bt)}X))(C\otimes I).    
\end{align}

\textit{Position Probability Dynamics}:
Selecting initial state $|00\rangle$ and  Hadamard gate as the coin operator $C$, the state evolves to 
\begin{align}
W(t)|00\rangle&=\frac{1}{\sqrt{2}}(|0\rangle(\cos{at}|0\rangle-i\sin{at}|1\rangle)\nonumber\\
&+|1\rangle(\cos{bt}|0\rangle-i\sin{bt}|1\rangle)).   
\end{align}
The probability  on vertex $1$ becomes $P_{1}(t)=\frac{1}{2}(\sin^2{(at)}+\sin^2{(bt)})=\frac{1}{2}(1-\cos{((a+b)t)}\cos{((a-b)t)})$.
The adjacency matrix of a 2-vertex circle is the Pauli $X$ matrix, thus the continuous quantum walk on the 2-vertex circle with the weight factor $\omega$ is $e^{-i\omega Xt}$. The probability of the walker position on vertex $1$ after $t$ time continuous quantum walk with initial state $|0\rangle$ is $\sin^2{(\omega t)}$. 

\begin{figure}[htpb]  
    \centering  
    \begin{subfigure}[htbp]{0.23\textwidth}  
        \centering  \includegraphics[width=\textwidth]{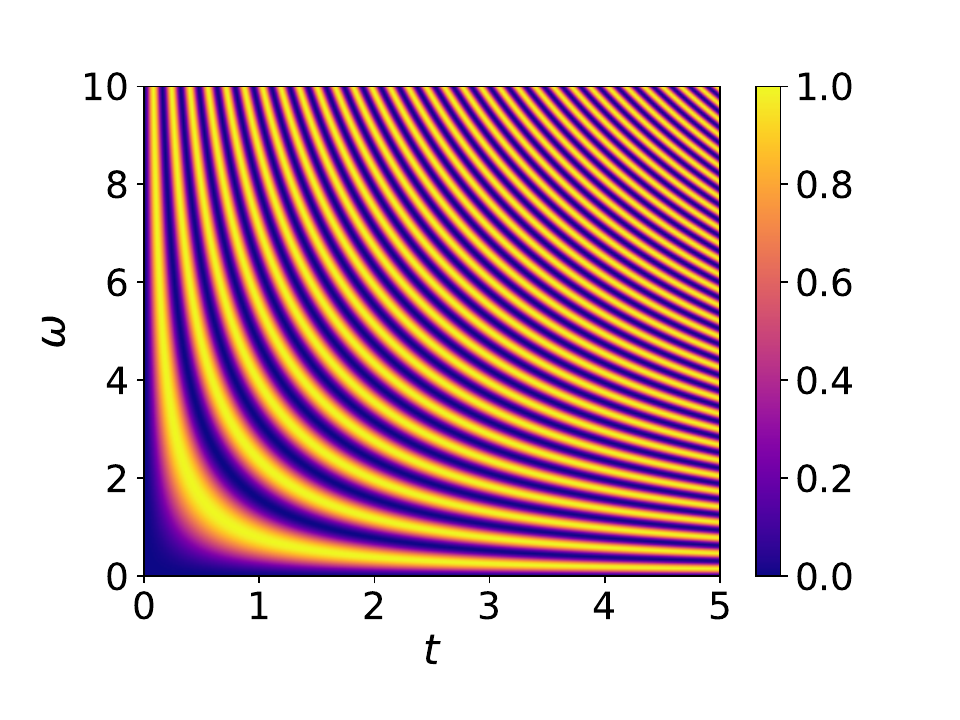}  
        \subcaption{}  
        \label{fig:2}  
    \end{subfigure}   
    \begin{subfigure}[htbp]{0.23\textwidth}  
        \centering          \includegraphics[width=\textwidth]{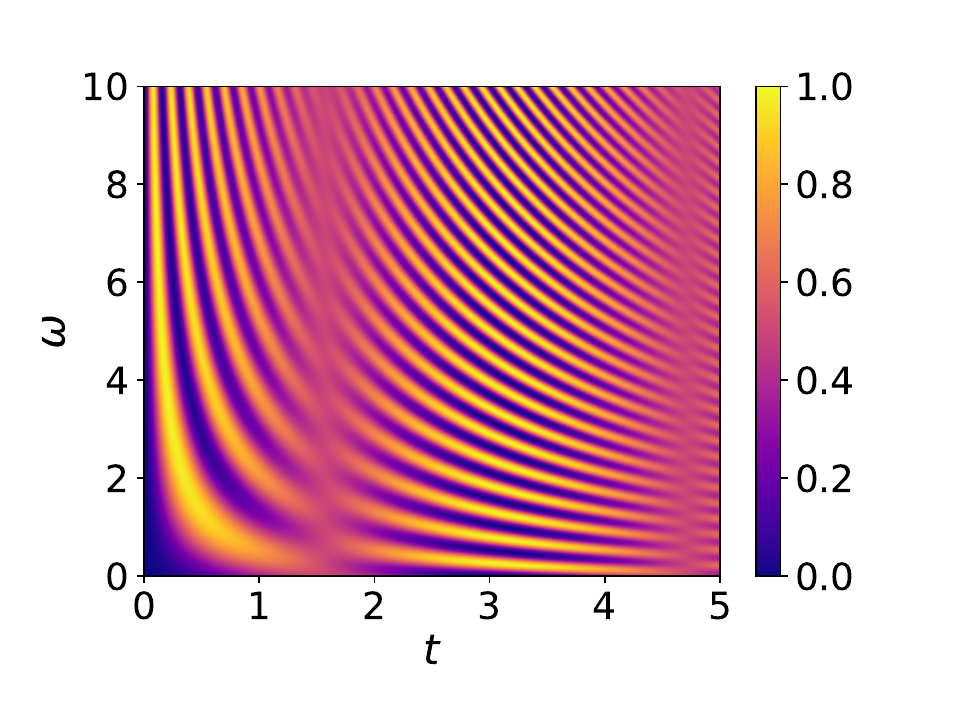}  
        \subcaption{}  
        \label{fig:3}  
    \end{subfigure}  

    \begin{subfigure}[htbp]{0.23\textwidth}  
        \centering  \includegraphics[width=\textwidth]{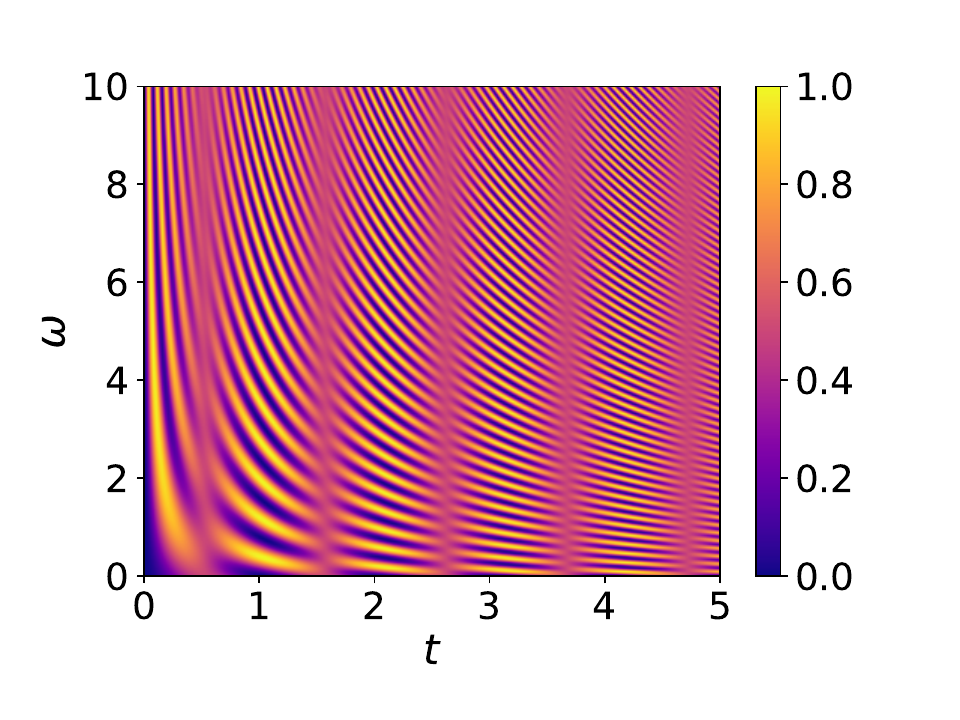}  
        \subcaption{}  
        \label{fig:4}  
    \end{subfigure}   
    \begin{subfigure}[htbp]{0.23\textwidth}  
        \centering          \includegraphics[width=\textwidth]{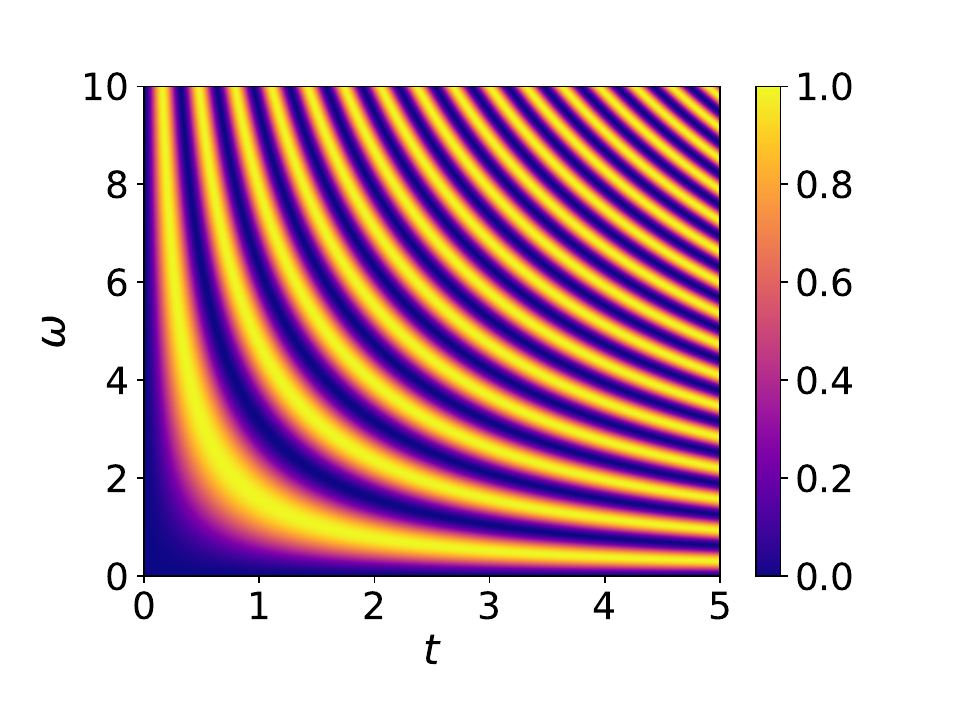}  
        \subcaption{}  
        \label{fig:5}  
    \end{subfigure} 
    
    \caption{The diagrams showing the probability of the position state being located at vertex 1 after a quantum walk on a 2-vertex circle graph starting from the initial state $|00\rangle$, varying with time $t$ and the weight factor $\omega$. (a)-(c) The hybrid quantum walks with different $a$ and $b$. (d) The continuous quantum walk.}  
    
    \label{fig:fig2}  
\end{figure}

FIG.~\ref{fig:fig2} illustrates the variation of probability $P_{1}(t)$ with respect to time $t$ and weight $\omega$. FIG.~\ref{fig:2}-FIG.~\ref{fig:4} also illustrate the impact of parameters $a$ and $b$ selection according to TABLE ~\ref{tab:1} on probability $P_{1}(t)$. From FIG.~\ref{fig:2} and FIG.~\ref{fig:5}, we can find when we select symmetry parameters, i.e. $|a-b|=0$, the hybrid quantum walk degenerates into continuous quantum walking. At this point, the probability $P_{1}(t)=\frac{1}{2}(1-\cos{(4\omega t)}$ only contains a single frequency $4\omega$. However, when $|a-b|\neq0$ and $|a-b|$ is independent of $\omega$, $P_{1}(t)=\frac{1}{2}$ at specific times $t=(\frac{\pi}{2}+k\pi)/|a-b|$, $k\in \mathbb{N}$. The breaking of parameter symmetry leads to special probability stable bands (pink stripes) in the system as shown in FIG.~\ref{fig:3} and FIG.~\ref{fig:4}. This is because asymmetric hybrid quantum walk breaks the uniformity limitation of continuous quantum walks by introducing phase parameters $a$ and $b$ that depend on coin states, making its dynamics dominated by dual frequency interference: high frequency oscillation $a+b$ and low frequency envelope $|a-b|$ revealed in $P_{1}(t)=\frac{1}{2}(1-\cos{((a+b)t)}\cos{((a-b)t)})$. High frequency oscillation term $\cos{((a+b)t)}$, controlling the oscillation frequency of the walker, is affected by low frequency envelope $\cos{((a-b)t)}$ which forms quantum beat frequency. Therefore, we can find that in FIG.~\ref{fig:3} and FIG.~\ref{fig:4}, as the high frequency $a+b$ increase, the oscillation frequency of the walker increases, that is, the yellow ripples become dense. And as the asymmetry of the two parameters increases from weak to strong, that is, $|a-b|$ increases, the beat frequency becomes larger, and the pink stripes appear more densely. This indicates that hybrid quantum walks can't only simulate continuous quantum walks, but also have unique dynamic characteristics such as dual frequency interference effect and quantum beat frequency effect. Its core innovation is to expand the single frequency dynamics of conventional continuous quantum walks into a multi-frequency controllable system, providing a new regulatory dimension for future quantum technology.

\textit{Universal Computation}: 
The quantum walk on a 2-vertex circle can achieve universal quantum gates. Since $S(0)=I\otimes I$, $W(0)=C\otimes I$ which is a single qubit unitary gate $C$ acting on the subspace $H_c$. If $\cos{at}=\pm 1, \sin{bt}=\pm1$, then $W(t)$ with a proper $C$ can become $|0\rangle\langle0|\otimes I+|1\rangle\langle1|\otimes X$ i.e. the CNOT gate. Thus to satisfy the above conditions, the range of values for $t$ should be the intersection of sets $\{\frac{k\pi}{a}: k\in\mathbb{N}\}$ and $\{\frac{\frac{\pi}{2}+k\pi}{b}: k\in\mathbb{N}\}$. Therefore, the intersection of the two sets will be non-empty only when $a$ and $b$ satisfy the condition that there exist $k, l\in\mathbb{N}$ such that $\frac{a}{b}=\frac{2k}{1+2l}$, in which case W can be a CNOT gate.

\subsection{N-star graph} 

\begin{figure}[htpb]  
    \centering  
    \begin{subfigure}[htbp]{0.46\textwidth}  
        \centering  \includegraphics[width=\textwidth]{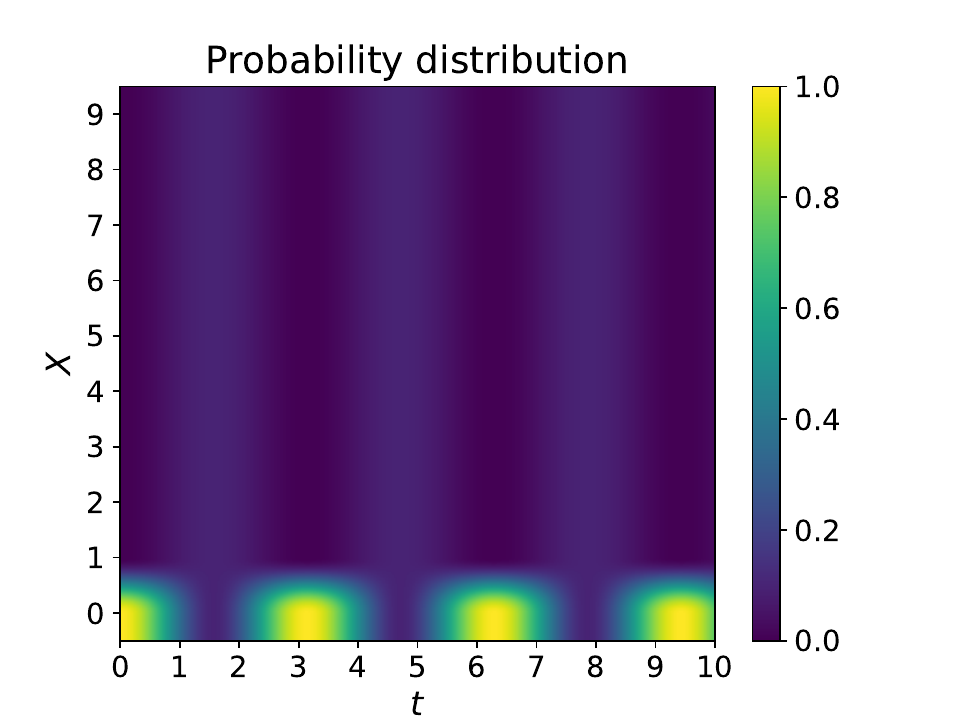}  
        \caption{}  
        \label{fig:6}  
    \end{subfigure}  
    
    \begin{subfigure}[htbp]{0.46\textwidth}  
        \centering          \includegraphics[width=\textwidth]{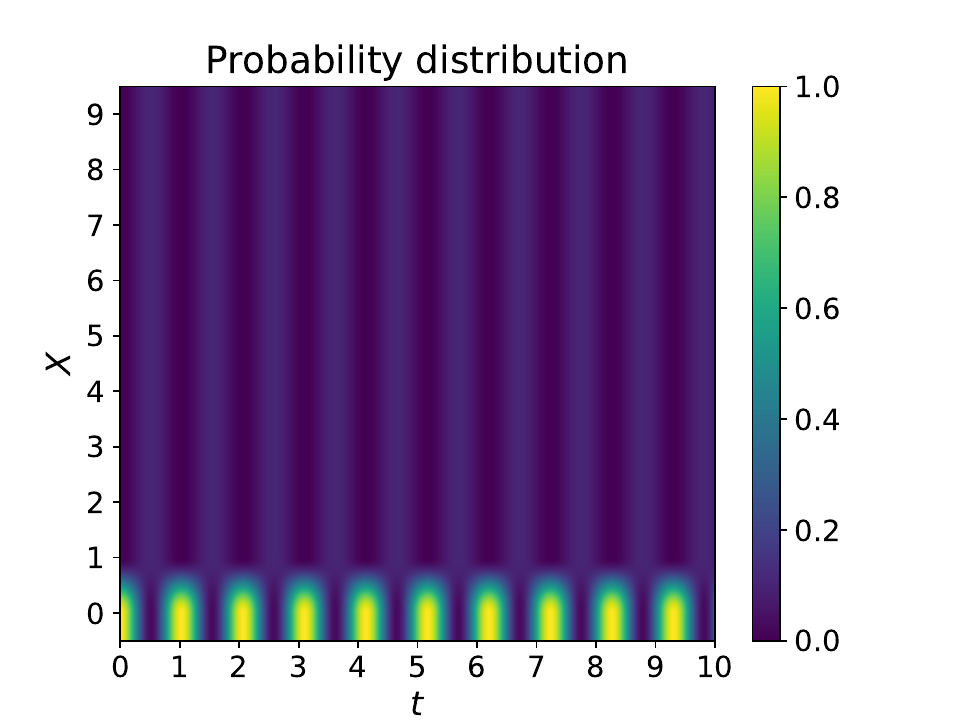} 
        \caption{}  
        \label{fig:7}  
    \end{subfigure}  

    \caption{Probability distributions of quantum walks on a 10-star graph vertex $X$, starting from the initial state $|00\rangle$, varying with time $t$. (a) The hybrid quantum walks. (b) The continuous quantum walk.}  
    \label{fig:fig3}  
\end{figure}

\begin{figure}[htbp]
  \centering
  \includegraphics[width=0.48\textwidth]{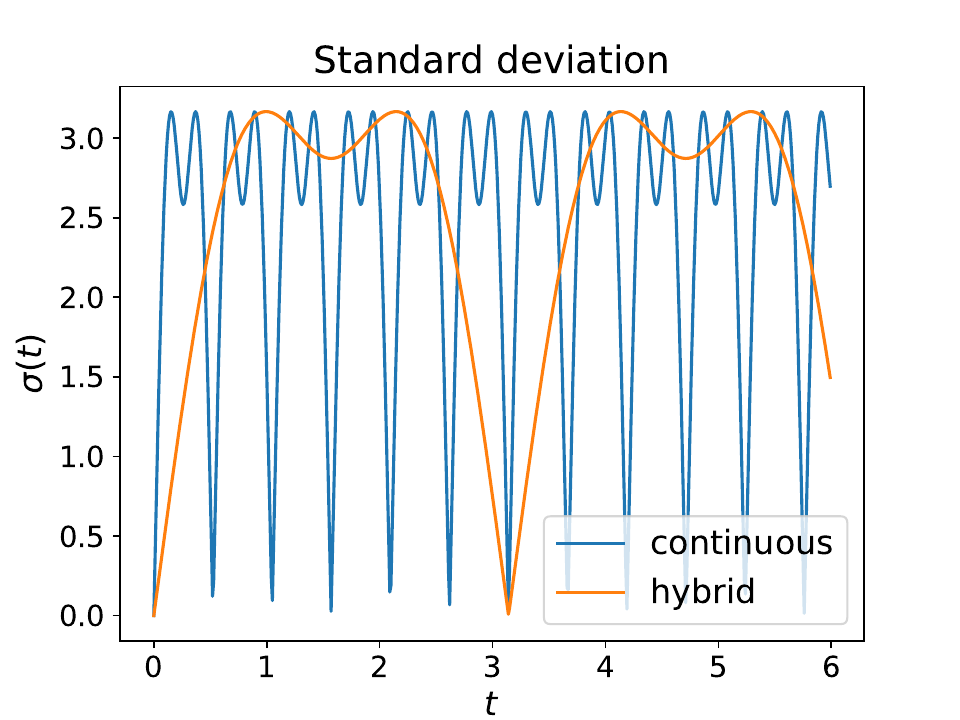}\\
  \caption{Standard deviations of the continuous quantum walk (blue line) and the hybrid quantum walk (orange line) on a 10-star graph, varying with time $t$.}
  \label{fig:8}
\end{figure}

Consider an $N$-vertex star graph $G=(V,E)$ with vertex set $V=\{0,1,\cdots,N-1\}$ and edge set $E=\{(0,j)\}_{j=1}^{N-1}.$ The composite Hilbert space has $N$ vertices and $N-1$ edges, in which a central vertex is connected to each of the other $N-1$ vertices by an edge. We denote $V$ as $\{0,1,\cdots,N-1\}$ and $\Gamma$ as $\{0,1,2,\cdots,N-1\}$, but the label $0$ without using. The walk space is $H_c\otimes H_p$, where $H_c=H_p=\mathrm{span}\{|0\rangle,\cdots,|N-1\rangle\}$.
The Hamiltonian is
\begin{align}
H=\sum_{j=1}^{N-1}|j\rangle\langle j|\otimes(|j\rangle\langle 0|+|0\rangle\langle j|).    
\end{align} 
The coin operator $C$ is the quantum Fourier operator. The initial state is $|00\rangle$ and the walk operator is $W(t)=e^{-iHt}(C\otimes I)$. After one step of the quantum walk with $t $ time, the quantum state becomes
\begin{align}
&|\phi(t)\rangle=W(t)|00\rangle\nonumber\\
&=\frac{1}{\sqrt{N}}(|00\rangle+\cos{t}\sum_{j=1}^{N-1}|j0\rangle-i\sin{t}\sum_{j=1}^{N-1}|jj\rangle).    
\end{align}
The probability of the walker position being on vertex $n$ is $P(n)=\frac{\sin^2t}{N}$, where $n=1,\cdots,N-1$, and $P(0)=\frac{(N-1)\cos^2t+1}{N}$. Thus the standard deviation is defined as  
\begin{align}
\sigma(t)=(\frac{\sin^2t}{N}\sum_{j=1}^{N-1}j^2-(\sum_{j=1}^{N-1}\frac{\sin^2t}{N}j)^2)^{1/2}.    
\end{align} 
The evolution of the continuous time quantum walk on the $N$-star graph is described by the unitary operator $U(t)=e^{-iAt}$, where $A=\sum_{j=1}^{N-1}(|j\rangle\langle 0|+|0\rangle\langle j|)$ is the adjacency matrix of the $N$-star graph. Then $U(t)|0\rangle=\cos{(\sqrt{N-1}t)}|0\rangle-\frac{i\sin{(\sqrt{N-1}t)}}{\sqrt{N-1}}\sum_{j=1}^{N-1}|j\rangle$. Thus the probability distribution and the standard deviation of the walker position are periodic functions with $\frac{\pi}{\sqrt{N-1}}$ and $\frac{\pi}{2\sqrt{N-1}}$ as their periods, respectively. But in our model, the probability distribution and the standard deviation is a $\pi$-periodic function without impacting from the vertex number $N$ as shown in FIG.~\ref{fig:fig3} and FIG.~\ref{fig:8}. This indicates that the position state is likely to be distributed at the endpoints, rather than the center point, with a high probability over a relatively long continuous period. This shows that our hybrid quantum walk is more parameter robust than the conventional continuous quantum walk: in the star graph of any size $N$, the evolution time parameter does not need to be adjusted due to $N$ changes, which can greatly simplify the experimental design and algorithm implementation. And our walk is more stable: the high probability distribution of the quantum state at the endpoint can be maintained for a long time, avoiding the frequent oscillation caused by the shortening of the period with $N$ in the traditional model. These characteristics make it have unique potential in quantum networks and other fields, especially for large-scale scalable star networks.   

\begin{figure}[htpb]  
    \centering  

    \begin{subfigure}[htbp]{0.46\textwidth}  
        \centering          \includegraphics[width=\textwidth]{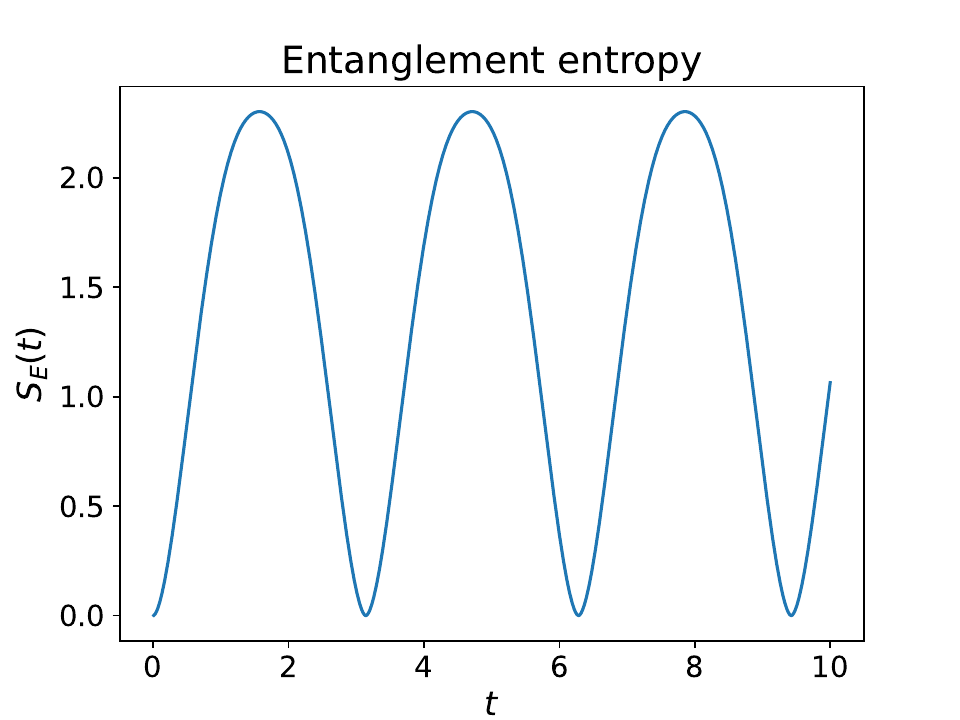} 
        \caption{}  
        \label{fig:9}  
    \end{subfigure} 
    
    \begin{subfigure}[htbp]{0.46\textwidth}  
        \centering          \includegraphics[width=\textwidth]{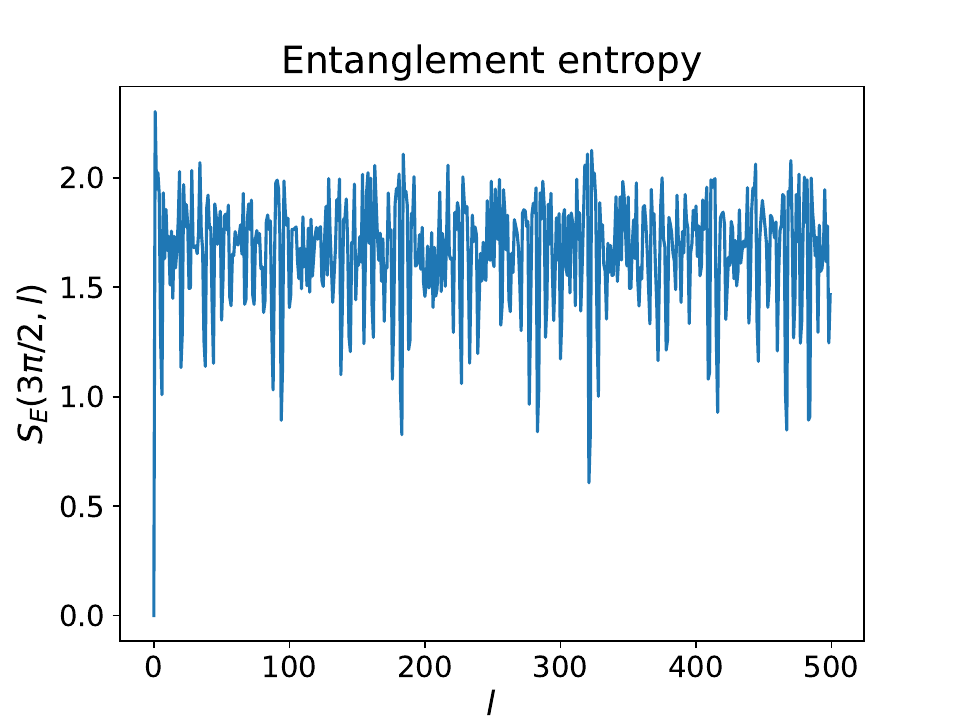} 
        \caption{}  
        \label{fig:10}  
    \end{subfigure} 
    \caption{The graphs of the entanglement entropy $S_E(t)$ and the entanglement entropy $S_E(\frac{3\pi}{2},l)$ on a 10-star graph.}  
    \label{fig:fig5}  
\end{figure}

The difference from continuous quantum walks lies in the fact that, due to the specific nature of the walk space and the framework of quantum walks with coins, we can naturally introduce entanglement into the quantum state after the evolution process. The density operator after a $t$ time hybrid quantum walk evolution is $\rho(t)=|\phi(t)\rangle\langle \phi(t)|$, thus the reduced density operator is $\rho_p(t)=$tr$_c(\rho(t))$. The entanglement entropy is defined as $S_E(t)=-$tr$(\rho_p(t)\log\rho_p(t))$. And the density operator after $l$ steps $t$ time hybrid quantum walk evolution is $\rho(t,l)=W(t)^{l}|\phi(0)\rangle\langle \phi(0)|(W(t)^{\dag})^l$, and the reduced density operator is $\rho_p(t,l)=$tr$_c(\rho(t,l))$, then the entanglement entropy is defined as $S_E(t,l)=-$tr$(\rho_p(t,l)\log\rho_p(t,l))$. In FIG.~\ref{fig:9} and FIG.~\ref{fig:10}, we show the function images of $S_E(t)$ and $S_E(\frac{3\pi}{2},l)$ on a 10-star graph. The value of $S_E(\frac{3\pi}{2},l)$ mostly falls within the range of [1, 2]. By utilizing the entangled structure of the quantum state $|\phi(t)\rangle$, we can set the evolution time to $\frac{\pi}{2}$, thereby obtaining high-dimensional Bell states, which are crucial resources for quantum communication. It is not difficult to see that $S_E(t)$ is also a $\pi$-periodic function independent of the vertex number $N$. This shows that the time control parameters (such as the optimal evolution time) of entanglement dynamics can be uniformly set regardless of the size of the star graph, which can be used to simplify the experimental realization of high-dimensional entanglement.   

\subsection{Line} 

\begin{figure}[htpb]  
    \centering  
    \begin{subfigure}[htbp]{0.46\textwidth}  
        \centering  \includegraphics[width=\textwidth]{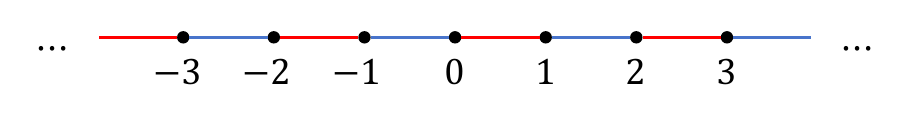}  
        \caption{}  
        \label{fig:24}  
    \end{subfigure}  
    
    \begin{subfigure}[htbp]{0.46\textwidth}  
        \centering          \includegraphics[width=\textwidth]{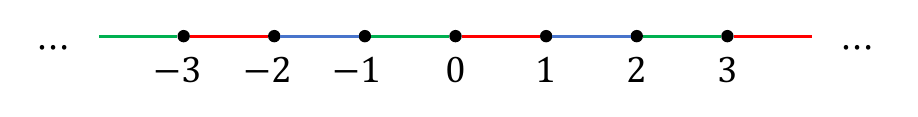} 
        \caption{}  
        \label{fig:25}  
    \end{subfigure}  

    \caption{Lines with some labels where red, blue, and green colors represent labels $0, 1, 2$, respectively. (a) The line with two labels. (b) The line with three labels.}  
    \label{fig:2425}  
\end{figure}

We consider the hybrid quantum walk on a line $G=(V, E)$, where $V=\mathbb{Z}$. We can label every edge of the line with some labels arbitrarily, but we focus on alternately labeling these edges with two or three labels to more specifically demonstrate the performance of the new model on a line. We use two labels to label these edges as shown in FIG.~\ref{fig:24}, so $\Gamma=\{0,1\}$ and $H_c\otimes H_p=\mathrm{span}\{|0\rangle,|1\rangle\}\otimes\mathrm{span}\{|x\rangle: x\in\mathbb{Z}\}$. The walk operator $W(t)=e^{-iHt}(C\otimes I)$, where
\begin{align}
H=|0\rangle\langle0|\otimes\sum_{x=2k,k\in\mathbb{Z}}(|x\rangle\langle x+1|+|x+1\rangle\langle x|)\nonumber\\
+|1\rangle\langle1|\otimes\sum_{x=2k+1,k\in\mathbb{Z}}(|x\rangle\langle x+1|+|x+1\rangle\langle x|),
\end{align} 
and we choose the Hadamard gate as the coin operator $C$. We also can use three labels to label these edges as shown in FIG.~\ref{fig:25}, so $\Gamma=\{0,1,2\}$ and $H_c\otimes H_p=\mathrm{span}\{|0\rangle,|1\rangle,|2\rangle\}\otimes\mathrm{span}\{|x\rangle: x\in\mathbb{Z}\}$. The walk operator $W(t)=e^{-iHt}\otimes(C\otimes I)$, where
\begin{align}
    H=|0\rangle\langle0|\otimes\sum_{x=3k,k\in\mathbb{Z}}(|x\rangle\langle x+1|+|x+1\rangle\langle x|)\nonumber\\+|1\rangle\langle1|\otimes\sum_{x=3k+1,k\in\mathbb{Z}}(|x\rangle\langle x+1|+|x+1\rangle\langle x|)\nonumber\\+|2\rangle\langle2|\otimes\sum_{x=3k+2,k\in\mathbb{Z}}(|x\rangle\langle x+1|+|x+1\rangle\langle x|)
\end{align} 
and we choose the 3-dim Grover operator as the coin operator $C$. We consider the hybrid quantum walks on a line with two or three labels and the classical discrete quantum walk and the continuous quantum walk $\exp\{-iHt\}$ on a line, where 
$H(i,j)=\frac{1}{\sqrt{2}}\delta_{ij}-\frac{1}{2\sqrt{2}}\delta_{i\pm1j}$, with the initial states being $\frac{|0\rangle-i|1\rangle}{\sqrt{2}}\otimes|0\rangle$, $\frac{|0\rangle+|1\rangle+|2\rangle}{\sqrt{3}}\otimes|0\rangle$,
$\frac{|0\rangle-i|1\rangle}{\sqrt{2}}\otimes|0\rangle$, and $|0\rangle$, respectively. The continuous evolution time $t$ of each step of the hybrid quantum walk is $\pi/2$. The step number $l$ is 100. 

\begin{figure}[htbp]
  \centering
  \includegraphics[width=0.48\textwidth]{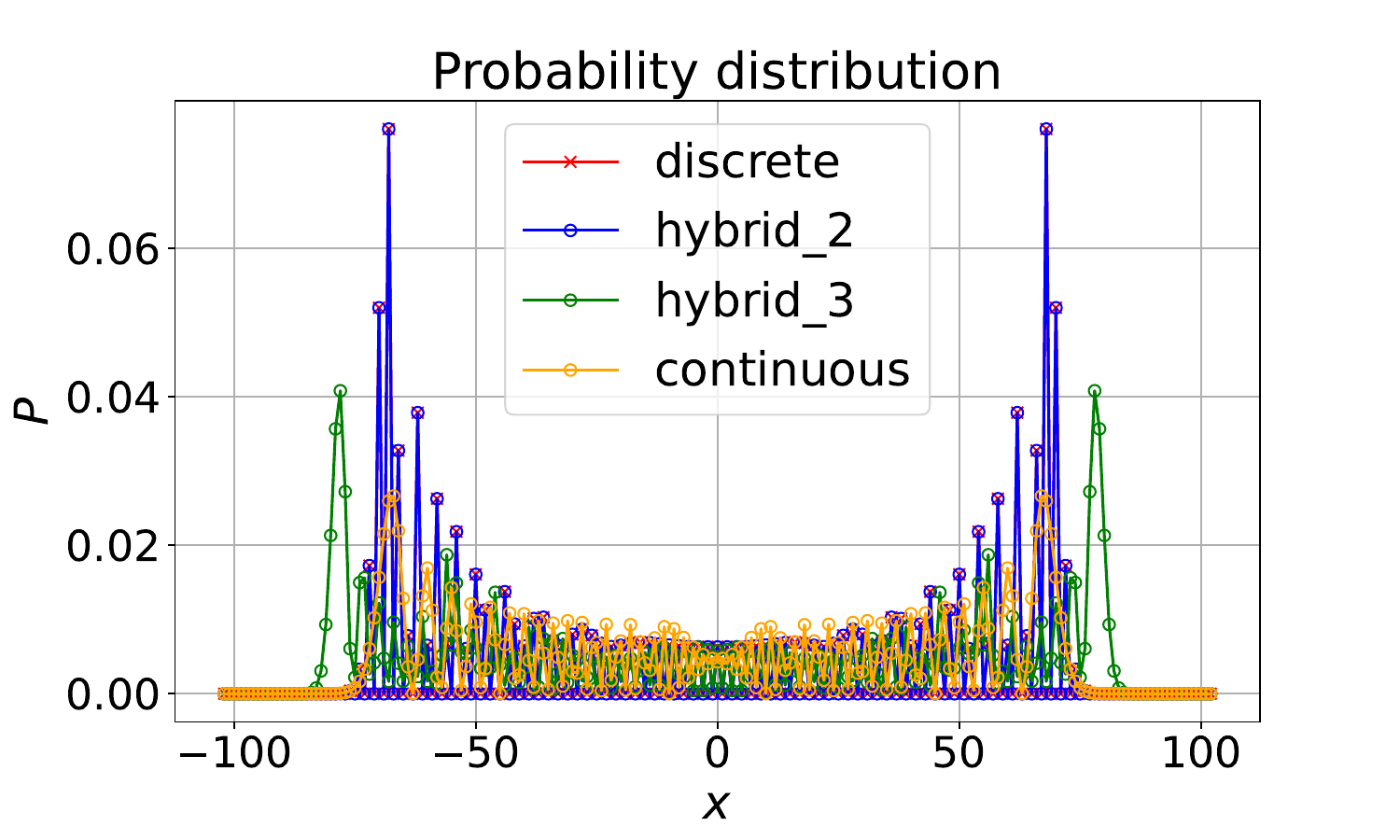}\\
  \caption{Probability distributions of quantum walks including the 100-step hybrid quantum walk on a line with two labels (blue dots), the 100-step hybrid quantum walk on a line with three labels (green dots), the classical 100-step discrete quantum walk (red crossings) and continuous quantum walk $\exp\{-100iH\}$ (orange dots).}
  \label{fig:11}
\end{figure}

For the hybrid quantum walks on a line with 2 labels, 
\begin{align}
    S(\frac{\pi}{2})=&-i|0\rangle\langle0|\otimes(\sum_{n\in odd}|n-1\rangle\langle n|+\sum_{n\in even}|n+1\rangle\langle n|)\nonumber\\
  &-i|1\rangle\langle1|\otimes(\sum_{n\in odd}|n+1\rangle\langle n|+\sum_{n\in even}|n-1\rangle\langle n|).
\end{align}
Now we transform the position space into the momentum space $\{|p\rangle: p\in[-\pi,\pi)\}$ through Fourier transformation to compare $W(\frac{\pi}{2})=S(\frac{\pi}{2})(H\otimes I)$ and the classical discrete quantum walk $S(H\otimes I)$, where 
\begin{align}
S=|0\rangle\langle0|\otimes\sum_{n\in \mathbb{Z}}|n+1\rangle\langle n|+|1\rangle\langle1|\otimes\sum_{n\in \mathbb{Z}}|n-1\rangle\langle n|.
\end{align} Through Fourier transformation, 
\begin{align}
    S(\frac{\pi}{2})=&-i|0\rangle\langle0|\otimes(\int_{-\pi}^{\pi}\int_{-\pi}^{\pi}\frac{\mathrm{d}p'\mathrm{d}p}{2\pi}(e^{-ip}\sum_{n\in odd}e^{i(p-p')n}\nonumber\\
    &+e^{ip}\sum_{n\in even}e^{i(p-p')n})|p\rangle\langle p'|\nonumber\\
  &-i|1\rangle\langle1|\otimes(\int_{-\pi}^{\pi}\int_{-\pi}^{\pi}\frac{\mathrm{d}p'\mathrm{d}p}{2\pi}(e^{ip}\sum_{n\in odd}e^{i(p-p')n}\nonumber\\&
    +e^{-ip}\sum_{n\in even}e^{i(p-p')n})|p\rangle\langle p'|.
\end{align}
Then we have 
\begin{align}
&S(\frac{\pi}{2})|+\rangle\otimes|p''\rangle
=\frac{|0\rangle}{\sqrt{2}}\otimes\int_{-\pi}^{\pi}\frac{\mathrm{d}p}{2\pi}(e^{-ip}\sum_{n\in odd}e^{i(p-p'')n}\nonumber\\
&+e^{ip}\sum_{n\in even}e^{i(p-p'')n})|p\rangle\nonumber\\
&+\frac{|1\rangle}{\sqrt{2}}\otimes\int_{-\pi}^{\pi}\frac{\mathrm{d}p}{2\pi}(e^{ip}\sum_{n\in odd}e^{i(p-p'')n}\nonumber\\
&+e^{-ip}\sum_{n\in even}e^{i(p-p'')n})|p\rangle.
\end{align}
Thus, 
\begin{align}
    &S(\frac{\pi}{2})|+\rangle\otimes|p\rangle=\cos{p}|+\rangle\otimes|p\rangle-i\sin{p}|-\rangle\otimes|p\rangle,\nonumber\\
    &S(\frac{\pi}{2})|-\rangle\otimes|p\rangle=-i\sin{p}|+\rangle\otimes|p\rangle+\cos{p}|-\rangle\otimes|p\rangle,
\end{align}
ignoring the global phase $-i$. 
And we have 
\begin{align}
&S|0\rangle\otimes|p\rangle=e^{-ip}|0\rangle\otimes|p\rangle,\nonumber\\
&S|1\rangle\otimes|p\rangle=e^{ip}|1\rangle\otimes|p\rangle \cite{26},
\end{align}
Thus, 
\begin{align}
&S|+\rangle\otimes|p\rangle=\cos{p}|+\rangle\otimes|p\rangle-i\sin{p}|-\rangle\otimes|p\rangle,\nonumber\\
    &S|-\rangle\otimes|p\rangle=-i\sin{p}|+\rangle\otimes|p\rangle+\cos{p}|-\rangle\otimes|p\rangle.
\end{align}
Therefore, we can obtain $W(\frac{\pi}{2})$ and $S(H\otimes I)$ have the same matrix representation under the basis $\{|\pm\rangle\otimes|p\rangle: p\in[-\pi, \pi)\}$, ignoring the global phase $-i$. As shown in FIG.~\ref{fig:11}, we can see that the evolution processions of the hybrid quantum walks on a line with 2 labels and the classical discrete quantum walk on a line are the same. Thus our model can simulate the classical discrete quantum walk on a line perfectly. This suggests that our hybrid model can, to a certain extent, incorporate the discrete quantum walk model, demonstrating strong generalization capabilities. 

For the hybrid quantum walk with $3$ labels, the line graph are divided into three non-intersecting sub-chains, and the evolution of each sub-chain is driven by the corresponding coin state. The effective Hamiltonian of $m$-th sub-chain is: $H_m=|m\rangle\langle m|\otimes\sum_{k\in\mathbb{Z}}(|3k+m\rangle\langle 3k+m+1|+|3k+m+1\rangle\langle 3k+m|$, $m=0,1,2$. This indicates that the dynamics of each sub-chain are independent of the others, forming three one-dimensional tightly bound chains, allowing transitions between adjacent positions only within each sub-chain. In order to analyze structures of block Hamiltonians and the dynamics of m-th sub-chains, it is necessary to transform the position space into the momentum space $P_m=\{|p\rangle_m, p\in[-\frac{\pi}{3},\frac{\pi}{3})\}$. The physical position interval of each sub-chain is 3, so the momentum $p\in[-\frac{\pi}{3},\frac{\pi}{3})$, and we define the following basis transformation:
\begin{align}
&|3k+m\rangle=\sqrt{\frac{3}{2\pi}}\int_{-\pi/3}^{\pi/3}\mathrm{d}pe^{ip(3k+m)}|p\rangle_m (*),\nonumber\\
&|p\rangle_m=\sqrt{\frac{3}{2\pi}}\sum_{k\in\mathbb{Z}}e^{-ip(3k+m)}|3k+m\rangle(**),
\end{align}
where $m=0,1,2$. We now need to prove that the above basis transformations are reciprocal. Substituting Formula (**) into the right-hand side of Formula (*), we obtain

\begin{align}
&\int_{-\pi/3}^{\pi/3} \frac{3\, \mathrm{d}p}{2\pi} e^{i p (3k + m)} \left( \sum_{k' \in \mathbb{Z}} e^{-i p (3k' + m)} |3k' + m\rangle \right) \nonumber \\
&= \sum_{k' \in \mathbb{Z}} \left( \int_{-\pi/3}^{\pi/3} \frac{3\, \mathrm{d}p}{2\pi} e^{i p [3(k - k')]} \right) |3k' + m\rangle.
\end{align}
After calculation, we have
\begin{equation}
    \int_{-\pi/3}^{\pi/3} \frac{3\, \mathrm{d}p}{2\pi} e^{i p [3(k - k')]} = \delta_{k,k'}.
\end{equation}
Thus, we can obtain the left-hand side of Formula (*): 
\begin{equation}
\sum_{k' \in \mathbb{Z}} \delta_{k,k'} |3k' + m\rangle =  |3k + m\rangle.
\end{equation}
Substituting Formula (*) into the right-hand side of Formula (**), we obtain
\begin{align}
 & \sum_{k \in \mathbb{Z}} e^{-i p (3k + m)} \left( \int_{-\pi/3}^{\pi/3} \frac{3\, \mathrm{d}p'}{2\pi} e^{i p' (3k + m)}  |p'\rangle_m \right) \nonumber \\
&= \int_{-\pi/3}^{\pi/3} \frac{3\, \mathrm{d}p'}{2\pi} \left( \sum_{k \in \mathbb{Z}} e^{i (p' - p)(3k + m)} \right) |p'\rangle_m.
\end{align}
According to the Poisson summation formula, we have
\begin{equation}
\sum_{k \in \mathbb{Z}} e^{i (p' - p)3k} = \frac{2\pi}{3} \sum_{n \in \mathbb{Z}} \delta\left(p' - p - \frac{2\pi n}{3}\right).
\end{equation}
Due to $p, p' \in [-\pi/3, \pi/3)$, only $n=0$ items contribute. Thus, we can obtain the left-hand side of Formula (**):
\begin{equation}
\int_{-\pi/3}^{\pi/3} \mathrm{d}p' \delta(p' - p)  |p'\rangle_m =  |p\rangle_m.
\end{equation}
So it is easy to calculate that the basis transformation defined above is well-defined.

So the Hamiltonian on the $m$-th sub-chain can be written in momentum representation as $H_m=|m\rangle\langle m|\otimes\hat{H}_m$, where $\hat{H}_m$ is
\begin{align}
\sum_{k\in\mathbb{Z}}\int_{-\frac{\pi}{3}}^{\frac{\pi}{3}} \int_{-\frac{\pi}{3}}^{\frac{\pi}{3}} \frac{3\mathrm{d}p\mathrm{d}p'}{2\pi}e^{-ip'}e^{i(3k+m)(p-p')}|p\rangle_{m,m+1}\langle p'|\nonumber\\
+\sum_{k\in\mathbb{Z}}\int_{-\frac{\pi}{3}}^{\frac{\pi}{3}} \int_{-\frac{\pi}{3}}^{\frac{\pi}{3}} \frac{3\mathrm{d}p\mathrm{d}p'}{2\pi}e^{ip'}e^{i(3k+m)(p'-p)}|p'\rangle_{m+1,m}\langle p|.  
\end{align}
And we have $_m\langle p|p'\rangle_{m'}=\delta(p-p')\delta_{mm'}$, so $\hat{H}_m|p''\rangle_{m+1}$ is
\begin{align}
&\sum_{k\in\mathbb{Z}}\int_{-\frac{\pi}{3}}^{\frac{\pi}{3}} \int_{-\frac{\pi}{3}}^{\frac{\pi}{3}} \frac{3\mathrm{d}p\mathrm{d}p'}{2\pi}e^{-ip'}e^{i(3k+m)(p-p')}\delta(p'-p'')|p\rangle_m\nonumber\\
&=e^{-ip''}\sum_{k\in\mathbb{Z}}\int_{-\pi/3}^{\pi/3} \frac{3\,\mathrm{d}p}{2\pi}e^{i(3k+m)(p-p'')}|p\rangle_m\nonumber\\
&=e^{-ip''}|p''\rangle_m. 
\end{align}
And $\hat{H}_m|p''\rangle_{m}=e^{ip''}|p''\rangle_{m+1}$. We set 
\begin{align}
|p_{\pm}\rangle_m=\frac{e^{-i\frac{p}{2}}|p\rangle_m\pm e^{i\frac{p}{2}}|p\rangle_{m+1}}{\sqrt{2}}.
\end{align}
Thus, we have 
\begin{align}
e^{-iHt}=\sum_{m=0}^2|m\rangle\langle m|\otimes e^{-i(|p_+\rangle_m\langle p_+|-|p_-\rangle_m\langle p_-|)t}.
\end{align}
Then we can get eigenvalues of $H$ are $E(p)=\pm1$, which indicates that the system has a flat energy band. Oscillation occurs only within the two-point line segment of the sub-chain, with no long-range propagation of energy or information. Therefore, coin operators and coin states are critical factors that significantly influence long-range diffusion. In the numerical simulation, we selected the 3-dim Grover operator as the coin operator, and the uniform superposition state as the initial state of the coin. And being similar to them, the probability distribution of the hybrid quantum walks on a line with three labels has two peaks. However, we can see that the probability distribution of the hybrid quantum walk with $3$ labels has two higher and further peaks than the continuous quantum walk, also further than the discrete quantum walk. It illustrates that choosing the appropriate coin operator makes the diffusion effect of our hybrid walk better than that of the conventional one. And the capability of our hybrid model can combine the unique strengths of both discrete and continuous quantum walk models, highlighting its robust fusion ability. In \textbf{Appendix A}, we also discuss various factors of initial coin states that influence the evolution process of our hybrid quantum walk on a line with three labels. 

\begin{figure}[htbp]
  \centering
  \includegraphics[width=0.47\textwidth]{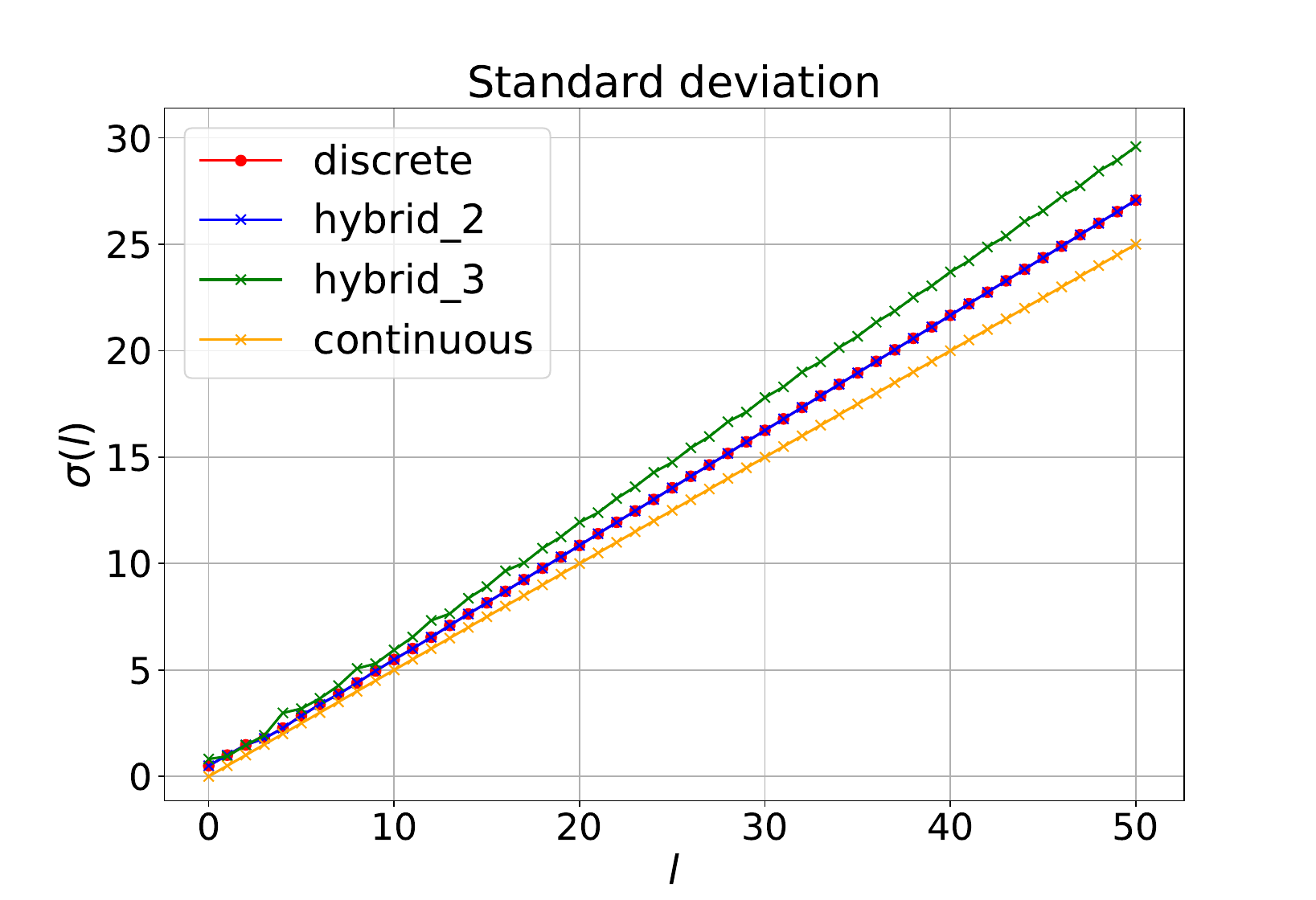}\\
  \caption{Standard deviations of quantum walks including the hybrid quantum walk on a line with two labels (blue crossings), the hybrid quantum walk on a line with three labels (green crossings), the classical discrete quantum walk (red dots), and the continuous quantum walk (orange crossings).}
  \label{fig:12}
\end{figure}

\begin{figure}[htbp]
  \centering
  \includegraphics[width=0.47\textwidth]{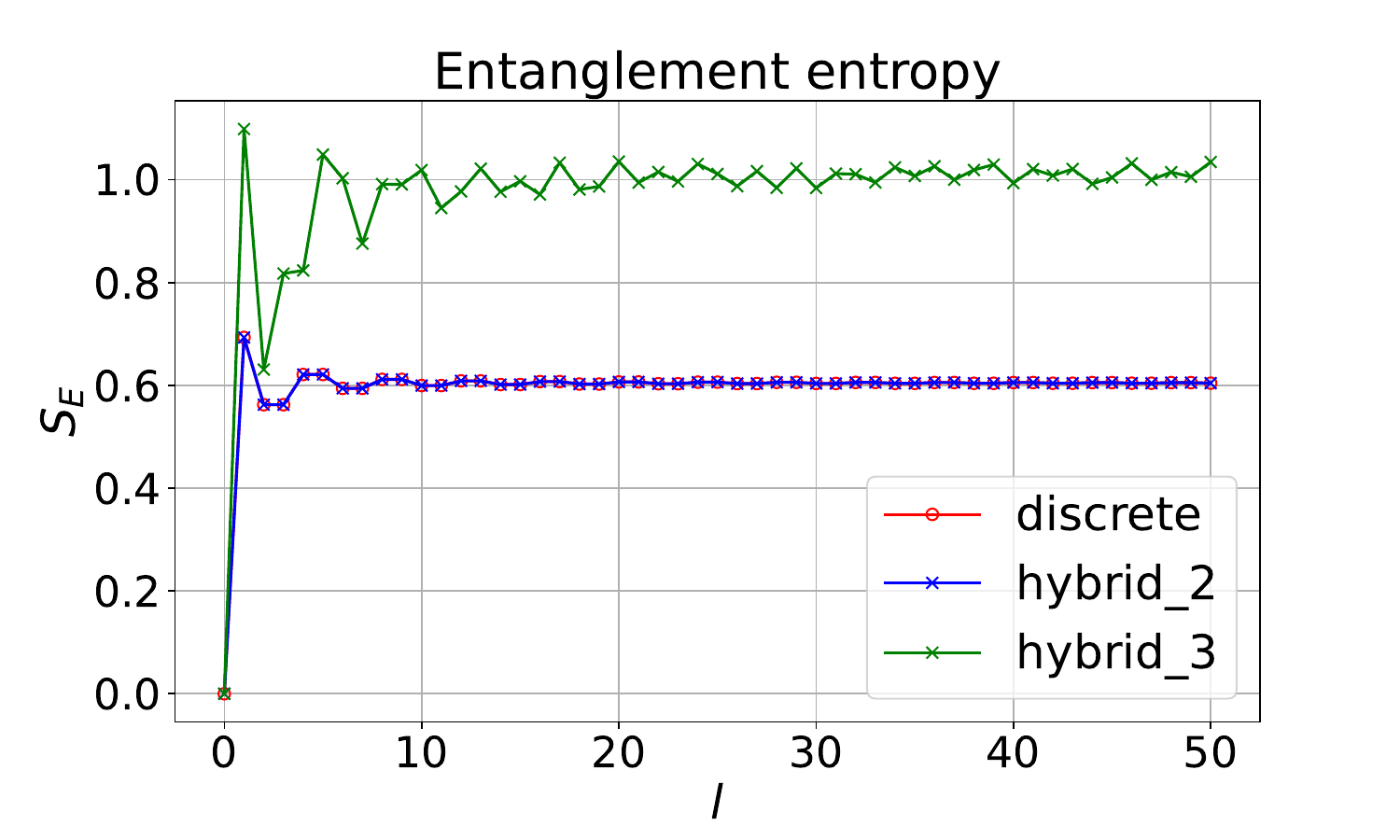}\\
  \caption{Entanglement entropy of quantum walks including the hybrid quantum walk on a line with two labels (blue crossings), the hybrid quantum walk on a line with three labels (green crossings), and the classical discrete quantum walk (red dots).}
  \label{fig:13}
\end{figure}

We calculate and show the function image of the standard deviations of the hybrid quantum walks on a line with two or three labels, the classical discrete and continuous quantum walk on a line as shown in FIG.~\ref{fig:12}, thus we obtain three lines with slopes being around 0.54, 0.59, 0.54, and 0.5. Due to the same evolution processions of the hybrid quantum walks on a line with two labels and the classical discrete quantum walk, their lines coincide naturally. It is evident that our hybrid quantum walk on a line with 3 labels achieves a higher standard deviation within the same time frame, indicating that its degree of diffusion is higher than that of other classical quantum walk models. This also shows that in our hybrid quantum walk, increasing the dimension of the coin space will increase the diffusion degree of the walker. Finally, we can find that using more labels leads to higher entanglement entropy from FIG.~\ref{fig:13}. Therefore, we can claim that increasing the coin space dimension by using more labels induces more entanglement resources. This will also provide new insights for the research on high-dimensional quantum state entanglement. 

\section{\MakeUppercase\expandafter{\romannumeral4}. Comparison with the discontinuous quantum walk}\label{s4}

\begin{figure}[htpb]  
    \centering  

    \begin{subfigure}[htbp]{0.46\textwidth}  
        \centering          \includegraphics[width=\textwidth]{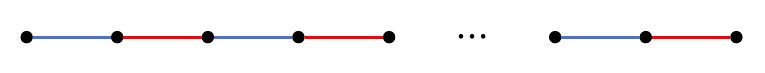} 
        \caption{}  
        \label{fig:14}  
    \end{subfigure} 
    
    \begin{subfigure}[htbp]{0.46\textwidth}  
        \centering          \includegraphics[width=\textwidth]{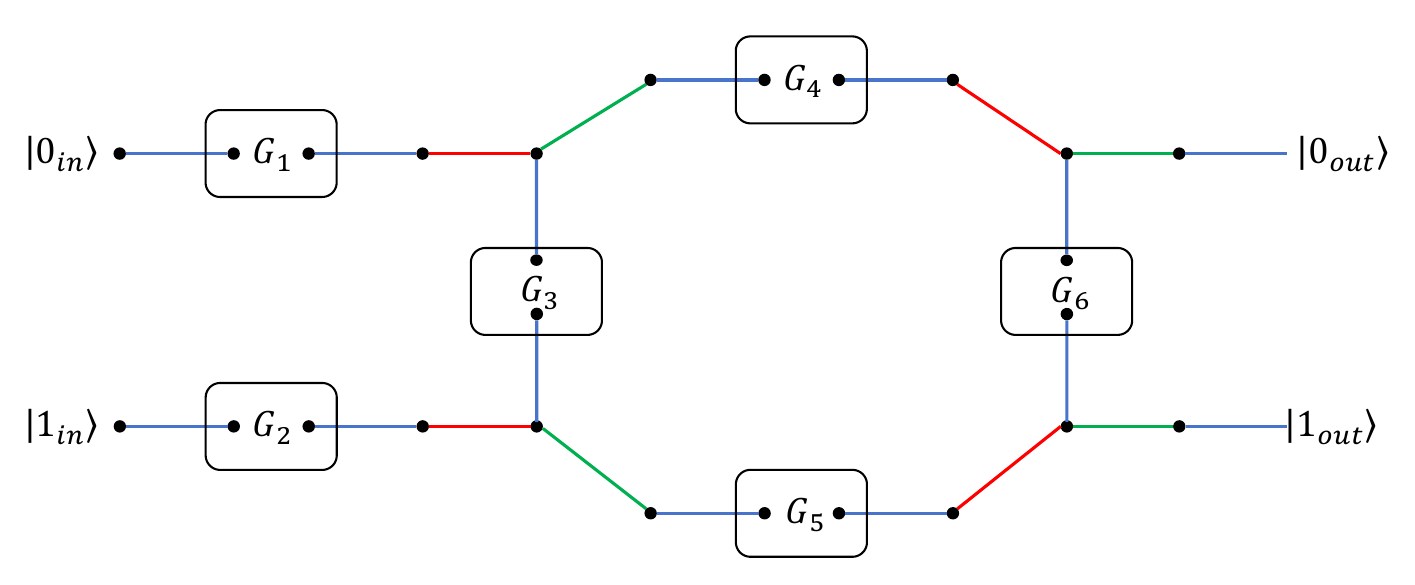} 
        \caption{}  
        \label{fig:15}  
    \end{subfigure} 
    \caption{(a) A line of some segments colored red and blue alternately. (b) An example of a single qubit operator implemented by quantum walks over the graph consisting of some red, blue green segments and some subgraph $G_j, j\in\mathrm{Z}$ \cite{3}.}  
    \label{fig:fig9}  
\end{figure}

The discontinuous quantum walk model proposed in \cite{3} implements universal quantum computation through perfect state transfer in continuous quantum walks on graphs by sequentially activating and deactivating graph segments. For a graph $G$, PST from vertex $a$ to vertex $b$ at time $t$ means that $|\langle b|e^{-iAt}|a\rangle|=1$, where $A$ is the adjacency matrix of $G$ and  $|a\rangle$, $|b\rangle$ are the quantum states corresponding to the vertices $a$ and $b$, respectively \cite{71}. While this approach achieves PST without coin operators, it requires precise dynamical control of edge weights in the Hamiltonian-a significant experimental challenge that \cite{3} does not fully address. Our hybrid quantum walk model provides a more practical alternative by replacing physical edge modulation with controlled coin operations in FIG.~\ref{fig:14}. The vertices of the line in FIG.~\ref{fig:14} are denoted as $\{1,\cdots,M\}$, and we label the red and blue segments as $r$ and $b$,respectively, thus the walk space $H_c\otimes H_p$ is $\mathrm{span}\{|r\rangle,|b\rangle\}\otimes\mathrm{span}\{|x\rangle:x\in\{1,\cdots,M\}\}$. The Hamiltonian of the quantum walk is $H=|r\rangle\langle r|\otimes S_r+|b\rangle\langle b|\otimes S_b$, where $S_r$ and $S_b$ are the adjacency matrices of the subgraphs generated by red segments and blue segments, respectively. The initial state is $|b\rangle|1\rangle$. After $M-1$ steps walks $W_{M-1}(\frac{\pi}{2})\cdots W_{1}(\frac{\pi}{2})$, where the coin operators $C_1=I$ and $C_i=|r\rangle\langle b|+|b\rangle\langle r|, i=2,\cdots,M-1$, the state becomes $|r\rangle|M\rangle$. Therefore, the state perfectly proceeds from position $1$ to position $M$. Obviously, the coin here plays a role in turning on and off, which enables the entire evolution process to be completed through a series of unitary operators in a closed quantum system. Our framework naturally extends to universal computation. For single-qubit operations in FIG.~\ref{fig:15}, (1) Algorithm-specific subgraphs $G_i (i\in\mathbb{Z})$ are selected; (2) Labeled segments (blue/red/green) define the Hamiltonian;(3) Adaptive coin operators route the state to $\alpha|0\rangle+\beta|1\rangle.$
The coin operators effectively emulate edge activation/deactivation while maintaining a closed quantum system evolution. So, different from the discontinuous model with external classical control operations, the model we propose can be more naturally interpreted as continuous-time quantum walks with internal degrees of freedom. This demonstrates our model's superior experimental feasibility and wider applicability compared to the discontinuous approach.
We now present additional applications of our hybrid quantum walk model.

\section{\MakeUppercase\expandafter{\romannumeral5}. Perfect state transfer on general connected graphs}\label{s5}

\begin{figure*}[htpb]  
\centering  

\begin{subfigure}[htbp]{0.45\textwidth}
\centering          
\includegraphics[width=1\textwidth]{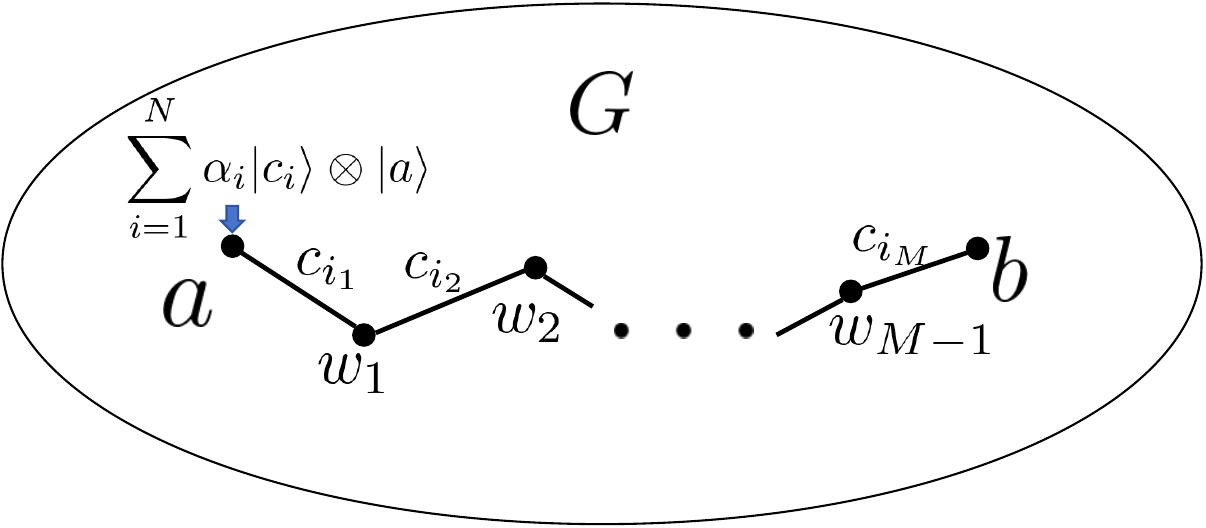} 
\caption{}  
\label{fig:29}  
\end{subfigure} 
\begin{subfigure}[htbp]{0.45\textwidth} 
\centering        
\includegraphics[width=1\textwidth]{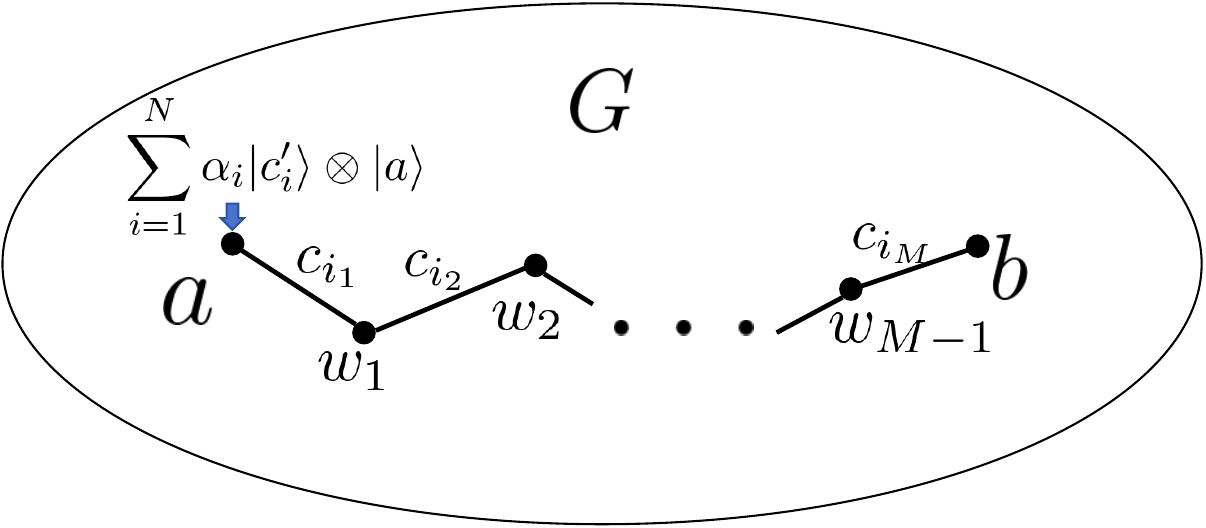} 
\caption{}  
\label{fig:30}  
\end{subfigure} 

\begin{subfigure}[htbp]{0.45\textwidth} 
\centering          
\includegraphics[width=1\textwidth]{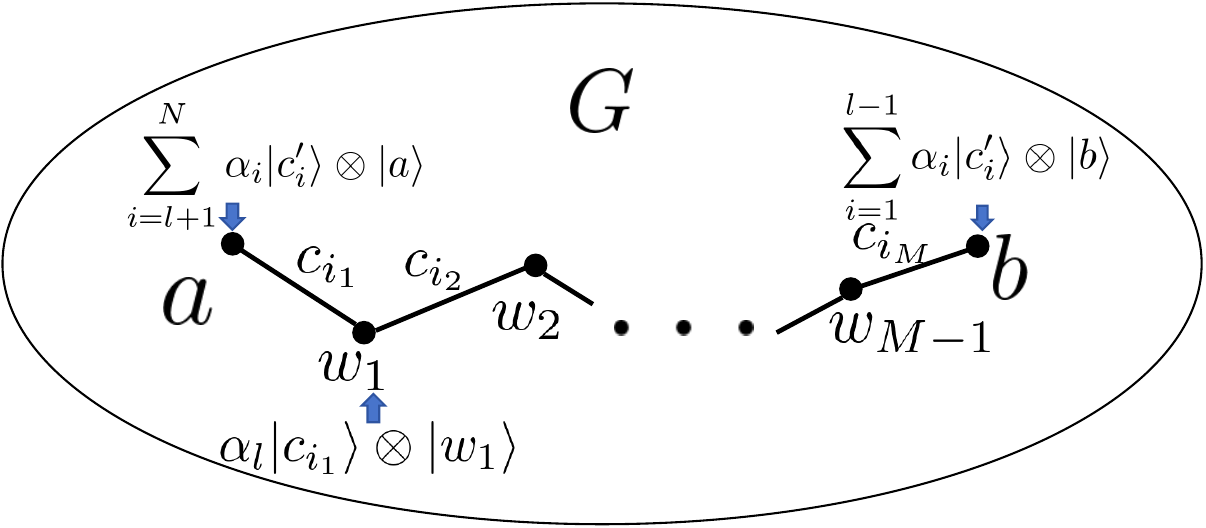} 
\caption{}  
\label{fig:31}  
\end{subfigure} 
\begin{subfigure}[htbp]{0.45\textwidth} 
\centering        
\includegraphics[width=1\textwidth]{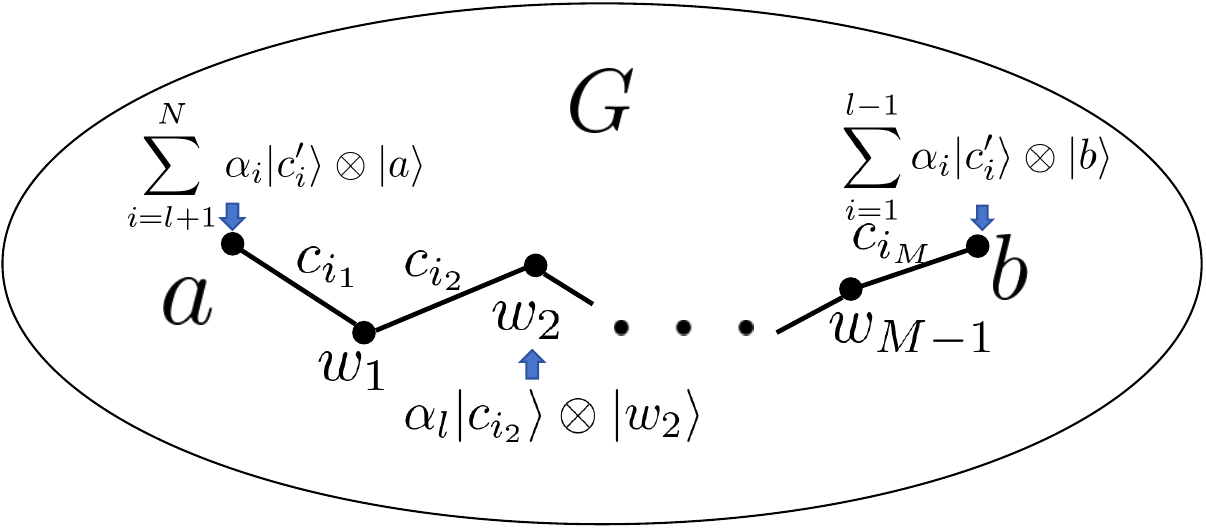} 
\caption{}  
\label{fig:32}  
\end{subfigure} 
\caption{(a) The path $aw_1\cdots w_{M-1}b$ in the graph $G$ and the initial state $\sum_{i=1}^{N}\alpha_i|c_i\rangle\otimes|a\rangle$. (b) The result of the operator $P$ acting on the initial state. (c) The result of the operator $e^{-iH\frac{3\pi}{2}}(D_l\otimes I)$ acting on the quantum state. The operator can only change the component $\alpha_l|c_l'\rangle\otimes|a\rangle$ to $\alpha_l|c_{i_1}\rangle\otimes|w_1\rangle$. (d) The result of the operator $e^{-iH\frac{3\pi}{2}}(C_1\otimes I)$ acting on the quantum state.} 
\label{fig:2932}  
\end{figure*}

Consider a general connected graph $G=(V, E)$ with a consistent coloring (adjacent edges have distinct colors) using the label set $\Gamma=\{c_1,\cdots,c_N\}$. For any two distinct vertices $ a, b\in V$ connected by a path $aw_1\cdots w_{M-1}b$ as shown in FIG.~\ref{fig:29}, we develop a protocol to achieve perfect state transfer from $a$ to $b$, i.e. the quantum state $\sum_{i=1}^{N}\alpha_i|c_i\rangle|a\rangle$ becomes $\sum_{i=1}^{N}\alpha_i|c_i\rangle|b\rangle$ after several steps quantum walks. 

Now, we set the quantum walk space and some operators. Firstly, we prepare a set $\Gamma'=\{c_1',\cdots,c_N'\}$ such that $\Gamma'\cap\Gamma=\emptyset$. Then we let the quantum walk space be $H_c\otimes H_p=\mathrm{span}\{|c\rangle:c\in\Gamma\cup\Gamma'\}\otimes\mathrm{span}\{|v\rangle:v\in V\}$. And $H=\sum_{c\in\Gamma\cup\Gamma'}|c\rangle\langle c|\otimes S_c$, 
where $S_c$ is the adjacent matrix of the graph $G_c=(V,E_c)$.
Next, we prepare some operators: 
\begin{align}
P=&\sum_{i=1}^{N}T_{c_i,c_i'}\otimes I,\nonumber\\
C_k=&T_{c_{i_{k+1}},c_{i_{k}}}+\sum_{\substack{i=1\\i\neq i_{k-1},i_k}}^{N}|c_i\rangle\langle c_i|+\sum_{i=1}^{N}|c_i'\rangle\langle c_i'|,\nonumber\\
D_k=&T_{c_{i_1},c_{k}'}+\sum_{\substack{i=1\\i\neq i_1}}^{N}|c_i\rangle\langle c_i|+\sum_{\substack{i=1\\i\neq k}}^{N}|c_i'\rangle\langle c_i'|,\nonumber\\
E_k=&T_{c_{i_M},c_{k}'}+\sum_{\substack{i=1\\i\neq i_M}}^{N}|c_i\rangle\langle c_i|+\sum_{\substack{i=1\\i\neq k}}^{N}|c_i'\rangle\langle c_i'|, k=1,\cdots,N,
\end{align}
where $T_{a,b}=|a\rangle\langle b|+|b\rangle\langle a|$.

We propose the perfect state transfer protocol as follows:
\begin{enumerate}
\item Perform the operator $P$ to the initial state $\sum_{i=1}^{N}\alpha_i|c_i\rangle\otimes|a\rangle$ as shown in FIG.~\ref{fig:29}, the state becomes $|\phi_0\rangle=\sum_{i=1}^{N}\alpha_i|c_i'\rangle\otimes|a\rangle$ as shown in FIG.~\ref{fig:30}. Altering the coin state of the initial state is to facilitate the control of each component of the quantum state during subsequent operations.  
\item Perform $N$ iterations of the quantum walks. In the $l^{th} (l=1,\cdots,N)$ iteration, we let the quantum walk operators be 
\begin{align}
&W_M(\frac{3\pi}{2})\cdots W_1(\frac{3\pi}{2})\nonumber\\
&=\prod_{k=M-1}^{1}(e^{-iH\frac{3\pi}{2}}(C_k\otimes I))e^{-iH\frac{3\pi}{2}}(D_l\otimes I).  
\end{align}
The quantum walk operator $e^{-iH\frac{3\pi}{2}}(D_l\otimes I)$ can change the component $\alpha_l|c_l'\rangle\otimes|a\rangle$ to $\alpha_l|c_{i_1}\rangle\otimes|w_1\rangle$ with the other components unchanging as shown in FIG.~\ref{fig:31}. Then $M-1$ steps of the hybrid quantum walks $\prod_{k=M-1}^{1}(e^{-iH\frac{3\pi}{2}}(C_k\otimes I))$ can transfer $\alpha_l|c_{i_1}\rangle\otimes|w_1\rangle$ to $\alpha_l|c_{i_M}\rangle\otimes|b\rangle$ with the other components unchanging as shown in FIG.~\ref{fig:32}.   
Then perform the operator $E_l$, thus the quantum state becomes 
\begin{align}
&|\phi_l\rangle=E_lW_M(\frac{3\pi}{2})\cdots W_1(\frac{3\pi}{2})|\phi_{l-1}\rangle\nonumber\\
&=(i)^M\sum_{i=1}^{l}\alpha_i|c_i'\rangle\otimes|b\rangle+\sum_{i=l+1}^{N}\alpha_i|c_i'\rangle\otimes|a\rangle. 
\end{align}
\item Perform the operator $P$ to the state $|\phi_N\rangle$, then the state $|\phi_N\rangle$ becomes $\sum_{i=1}^{N}\alpha_i|c_i\rangle\otimes|b\rangle$ which accomplishes the perfect state transfer. 
\end{enumerate}
The specific algorithm process is described in \textbf{Algorithm 1}.  

\begin{algorithm}[H]
\caption{Perfect State Transfer Protocol}
\label{alg:pst}
\begin{algorithmic}[1]
\Require Initial state $\sum_{i=1}^{N}\alpha_i|c_i\rangle\otimes|a\rangle$
\Ensure Final state $\sum_{i=1}^{N}\alpha_i|c_i\rangle\otimes|b\rangle$

\State \textbf{Step 1: Preparation}
\State Apply operator $P = \sum_{i=1}^{N}T_{c_i,c_i'}\otimes I$ 
\State Initial state becomes $|\phi_0\rangle = \sum_{i=1}^{N}\alpha_i|c_i'\rangle\otimes|a\rangle$ 

\State \textbf{Step 2: Quantum Walk Iterations}
\For{$l = 1$ \textbf{to} $N$}
    \State Define quantum walk operators:
    \[W_M(\tfrac{3\pi}{2})\cdots W_1(\tfrac{3\pi}{2})\]
    \[= \prod_{k=M-1}^{1}\left(e^{-iH\tfrac{3\pi}{2}}(C_k\otimes I)\right)e^{-iH\tfrac{3\pi}{2}}(D_l\otimes I)
    \]
    
    \State Execute operations:
    \begin{itemize}
        \item Transform component $\alpha_l|c_l'\rangle\otimes|a\rangle$ to $\alpha_l|c_{i_1}\rangle\otimes|w_1\rangle$ via $e^{-iH\tfrac{3\pi}{2}}(D_l\otimes I)$ (other components unchanged)
        \item Perform $M-1$ hybrid quantum walk steps $\prod_{k=M-1}^{1}\left(e^{-iH\tfrac{3\pi}{2}}(C_k\otimes I)\right)$ to transfer $\alpha_l|c_{i_1}\rangle\otimes|w_1\rangle$ to $\alpha_l|c_{i_M}\rangle\otimes|b\rangle$
    \end{itemize}
    
    \State Apply operator $E_l$ to obtain:
    \[
    |\phi_l\rangle = (i)^M\sum_{i=1}^{l}\alpha_i|c_i'\rangle\otimes|b\rangle + \sum_{i=l+1}^{N}\alpha_i|c_i'\rangle\otimes|a\rangle
    \]
\EndFor

\State \textbf{Step 3: Final Transformation}
\State Apply operator $P$ to transform $|\phi_N\rangle$ into $\sum_{i=1}^{N}\alpha_i|c_i\rangle\otimes|b\rangle$
\end{algorithmic}
\end{algorithm}

According to the above state transfer protocol, coin operators play the role of pointing, enabling the continuous evolution of each stage to proceed along the given path. Not only that, coin operators also play a regulating role. After expanding the dimension of the coin state, each component of the state can be transferred in turn. This demonstrates that the combination of the coin operator's regulation over the evolution process and the continuous evolution is crucial for implementing quantum algorithms. Because the conventional quantum walk with coins on the graph directly regards the process of controlled position state transfer as a conditional transfer operator, in order to keep the operator must be unitary, the selection of graphs is limited. However, our hybrid quantum walks treat the controlled transfer process of position states as a Hamiltonian of a continuous evolution, so we only need to ensure the Hermitian property of the Hamiltonian, which greatly relaxes the choice of graphs. Therefore, compared to previous schemes that utilized quantum walks for perfect state transfer on specific types of graphs \cite{21}, our scheme for perfect state transfer is suitable for general connected graphs. However, our scheme can be simplified according to the specific structure of the graph, so we only provide the most general approach here.

\begin{figure}[htpb]  
    \centering  

    \begin{subfigure}[htbp]{0.46\textwidth}  
        \centering          \includegraphics[width=\textwidth]{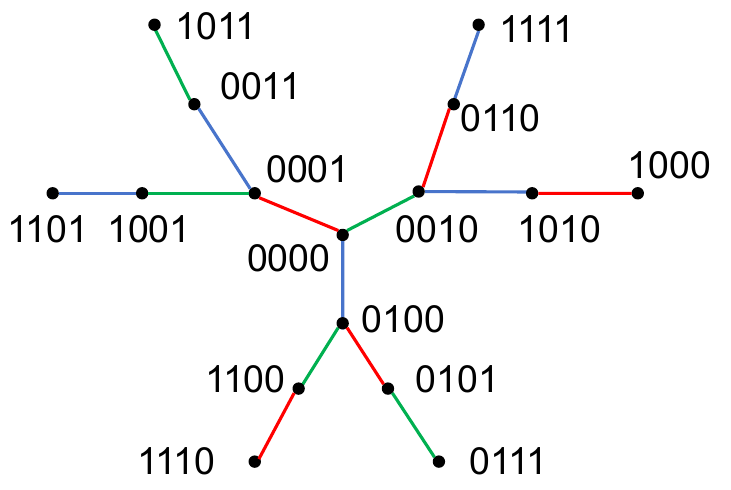} 
        \caption{}  
        \label{fig:33}  
    \end{subfigure} 
    
    \begin{subfigure}[htbp]{0.46\textwidth}  
        \centering          \includegraphics[width=\textwidth]{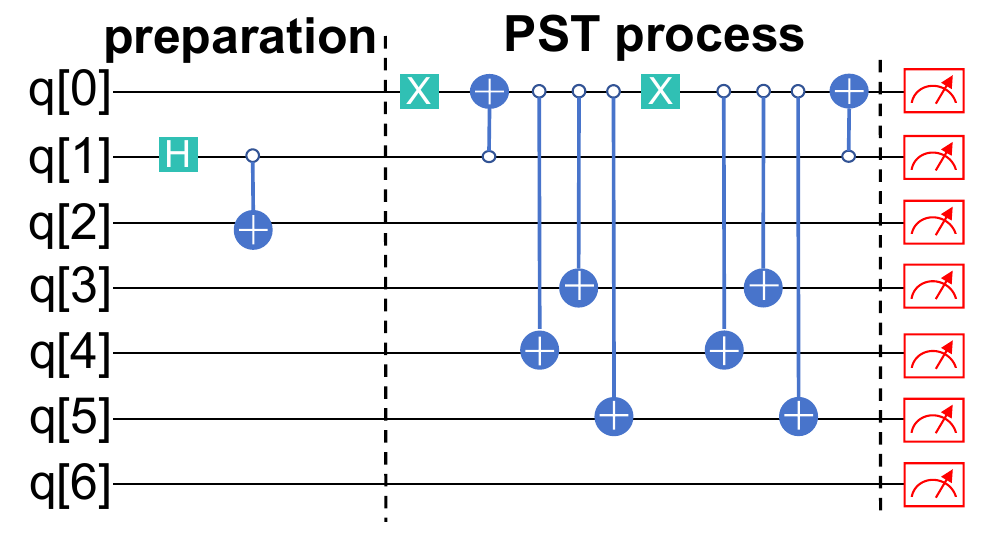} 
        \caption{}  
        \label{fig:34}  
    \end{subfigure} 
    \caption{(a) The tree diagram for implementing PST.The red, green and blue edge labels are 0, 1, and 2, respectively. (b) The simplified quantum circuit for implementing the PST process.}  
    \label{fig:fig15}  
\end{figure}

\begin{figure}[htpb]  
    \centering  

    \begin{subfigure}[htbp]{0.46\textwidth}  
        \centering          \includegraphics[width=\textwidth]{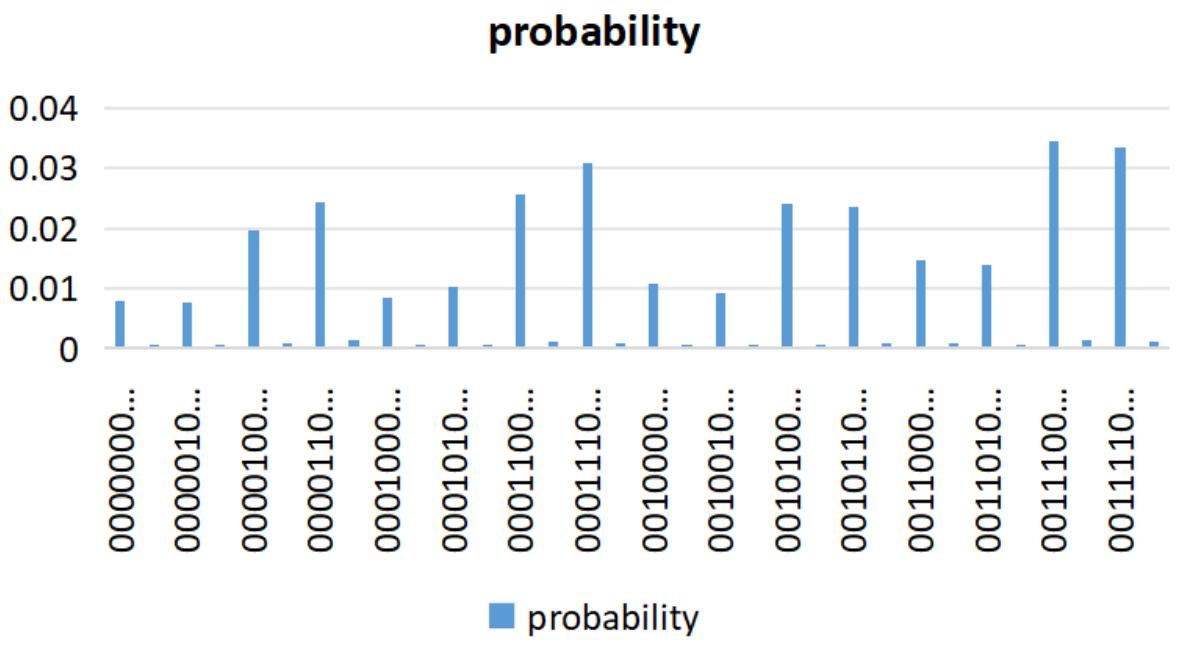} 
        \caption{}  
        \label{fig:35}  
    \end{subfigure} 
    
    \begin{subfigure}[htbp]{0.46\textwidth}  
        \centering          \includegraphics[width=\textwidth]{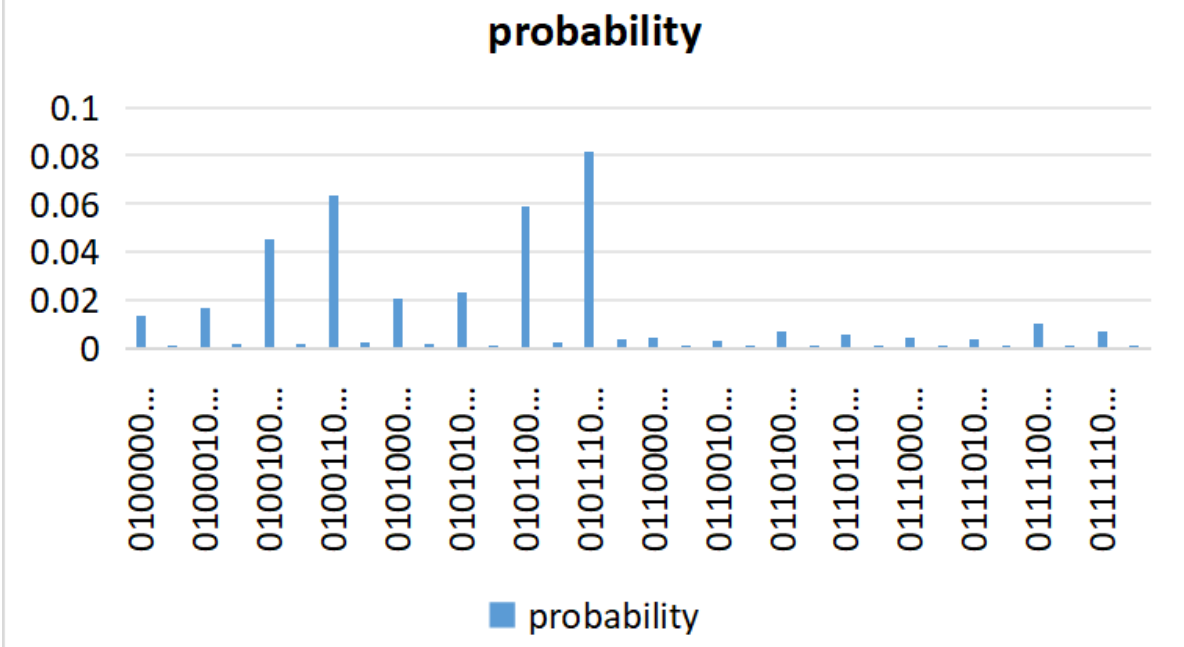} 
        \caption{}  
        \label{fig:36}  
    \end{subfigure} 
    
    \begin{subfigure}[htbp]{0.46\textwidth}  
        \centering          \includegraphics[width=\textwidth]{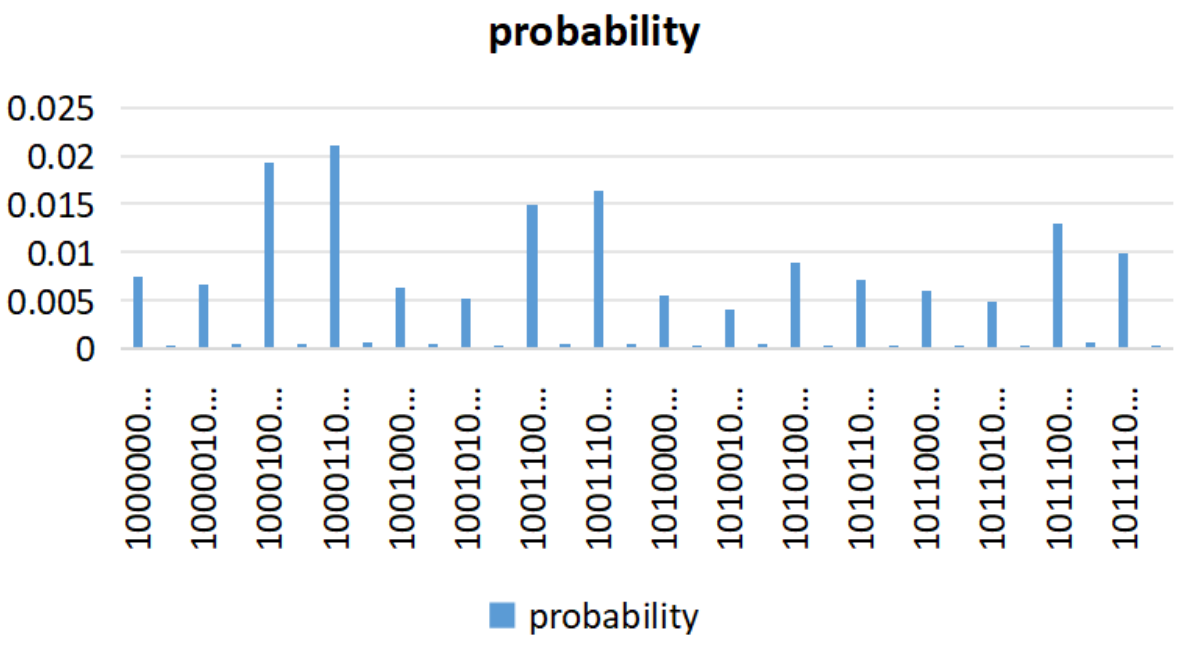} 
        \caption{}  
        \label{fig:37}  
    \end{subfigure} 
    
    \begin{subfigure}[htbp]{0.46\textwidth}  
        \centering          \includegraphics[width=\textwidth]{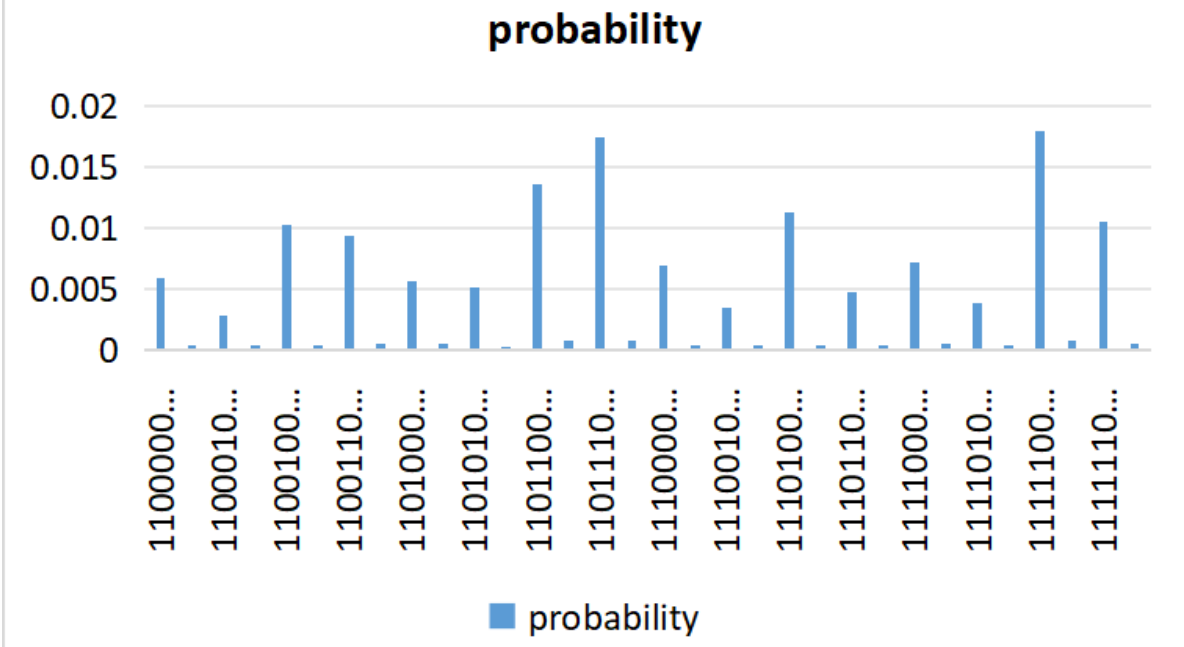} 
        \caption{}  
        \label{fig:38}  
    \end{subfigure} 
    \caption{(a)-(b) The probability values of all measurement results.}  
    \label{fig:fig16}  
\end{figure}

To demonstrate that our protocol can be applied to general connected graphs, we consider the experiment implementation of the PST on FIG.~\ref{fig:33} in a superconducting quantum processor. We can use the above protocol to achieve a state transfer of coin state $\frac{|001\rangle+|010\rangle}{\sqrt{2}}$ from vertex $0000$ to vertex $1110$, i.e. the quantum state $\frac{|001\rangle+|010\rangle}{\sqrt{2}}|0000\rangle$ becomes $\frac{|001\rangle+|010\rangle}{\sqrt{2}}|1110\rangle$. However, in order to greatly reduce the number of quantum gates in the experimental circuit, we reduce the redundant operations that do not affect the final output, such as some operations on coin states, and simplify the hybrid quantum walk operator according to the practical example. We finally obtain the quantum circuit in FIG.~\ref{fig:34}. The shots in our experiment are selected as 20000. The final measurement results are shown in FIG.~\ref{fig:fig16}. Although the measurement results are quite different from the target state due to quantum circuit errors, in the order of probability values of all measurement results from largest to smallest, the probabilities of the measurement results being 0011110 and 0101110 are 0.0335 and 0.0816, being the third and first, respectively.

\section{\MakeUppercase\expandafter{\romannumeral6}. matrix multiplication and triangle counting of regular graphs}\label{s6}

The matrix multiplication algorithm plays a crucial role in multiple fields and serves as a fundamental tool for solving various problems. The complexity of the fastest matrix multiplication algorithm currently stands at $O(n^{2.371552})$ \cite{49}, where $n$ is the dimension of the matrix. 

Recent advances in quantum matrix multiplication highlight distinct algorithmic strategies. \cite{54} introduced three approaches using the SWAP test, SVE, and HHL, with the SWAP test achieving the optimal complexity of $\Tilde{O}\big(\frac{n^2}{\varepsilon}\big)$ for precision $\varepsilon$, making it the primary candidate. In contrast, \cite{55} encodes matrices into basis states with $O(n^2\log n)$ complexity but outputs only a superposition of the product, limiting its utility to non-measurement intermediate processes. Meanwhile, \cite{56} employs block encoding with complexity $O\big(\frac{n^2}{\varepsilon^2}\big)$, relying on QRAM \cite{57} for state preparation. For the output results, \cite{55} stores results in basis states, whereas \cite{54} and \cite{56} encode products into quantum state amplitudes. For the latter, the element $C_{ij}$ of product matrix $C=AB$ satisfies 
\begin{align}
C_{ij} = \|A\|_F\|B\|_F \cdot \langle i,j|\langle 0|\psi_f\rangle,
\end{align}
where $\|A\|_F$ denotes the Frobenius norm. To estimate $C_{ij}$ within error $\delta$, the amplitude estimation precision must satisfy $\varepsilon \leq \delta/(\|A\|_F\|B\|_F)$, implying $\varepsilon^{-1}$ scales linearly with $\|A\|_F\|B\|_F$, thereby linking resource costs directly to input matrix norms.

In this section, we propose a quantum algorithm for computing the product of adjacency matrices of regular graphs by leveraging hybrid quantum walks and generalized CNOT gates. For example, PageRank \cite{58} calculation and non-negative matrix factorization algorithms \cite{59} require the computation of adjacency matrix products. For $d_l$-regular $n$ vertices graph $G_l=(V_l,E_l) (l=1, \cdots, K)$, the adjacency matrices are denoted as $A^{(l)}$. If $(i,j)\in E_l$ $A^{(l)}_{ij}=1$, otherwise $A^{(l)}_{ij}=0$. We let $\Gamma=\{0,1,\cdots,n-1\}$ be the edge label. For the hybrid quantum walk on the graph $G_l$, the walk space is $H_c\otimes H_p=\mathrm{span}\{|k\rangle: k \in \Gamma\}\otimes \mathrm{span}\{|k\rangle: k \in V_l\}$. And the Hamiltonian is 
\begin{align}
H_l&=\sum_{k=0}^{n-1}|k\rangle\langle k|\otimes S_k^{(l)}\nonumber\\
&=\sum_{k=0}^{n-1}|k\rangle\langle k|\otimes(\sum_{j=0}^{n-1}A^{(l)}_{jk}(|j\rangle\langle k|+|k\rangle\langle j|)).  
\end{align}
The coin operator is the identity operator. Thus the hybrid quantum walk operator $W^{(l)}(t)$ on the graph $G_l$ is
\begin{align}
\exp{\left(-i\left(\sum_{k=0}^{n-1}|k\rangle\langle k|\otimes S_k^{(l)}\right)t\right)}=\sum_{k=0}^{n-1}|k\rangle\langle k|\otimes e^{-iS_k^{(l)}t}.   
\end{align}
When $d_l \geq 2$, the rank of $S_k^{(l)}$ is $2$. Therefore, $S_k^{(l)}$ only has two nonzero eigenvalues, $\lambda_{k,\pm}^{(l)}=\pm\sqrt{d_l}$, with corresponding eigenvectors $|\lambda_{k,\pm}^{(l)}\rangle$ are
\begin{align}
(\frac{A_{0k}^{(l)}}{\sqrt{2d_l}},\cdots,\frac{A_{(k-1)k}^{(l)}}{\sqrt{2d_l}},\pm\frac{1}{\sqrt{2}},\frac{A_{(k+1)k}^{(l)}}{\sqrt{2d_l}},\cdots,\frac{A_{(n-1)k}^{(l)}}{\sqrt{2d_l}})^T.  
\end{align}
Thus $S_k^{(l)}=\lambda_{k,+}^{(l)}|\lambda_{k,+}^{(l)}\rangle\langle\lambda_{k,+}^{(l)}|+\lambda_{k,-}^{(l)}|\lambda_{k,-}^{(l)}\rangle\langle\lambda_{k,-}^{(l)}|$. Furthermore, the walk operator $W^{(l)}(t)$ is
\begin{align}
&\sum_{k=0}^{n-1}|k\rangle\langle k|\otimes (e^{-i\lambda_{k,+}^{(l)}t}|\lambda_{k,+}^{(l)}\rangle\langle\lambda_{k,+}^{(l)}|+e^{-i\lambda_{k,-}^{(l)}t}|\lambda_{k,-}^{(l)}\rangle\langle\lambda_{k,-}^{(l)}|\nonumber\\
&+I-|\lambda_{k,+}^{(l)}\rangle\langle\lambda_{k,+}^{(l)}|-|\lambda_{k,-}^{(l)}\rangle\langle\lambda_{k,-}^{(l)}|).
\end{align}
We denote the generalized CNOT operator 
\begin{align}
\sum_{i,j=0}^{n-1}|i\rangle_{control}\langle i|\otimes|(j+i)\mathrm{mod} n\rangle_{target}\langle j|   
\end{align}
as $GC_{control,target}$. Using the quantum adder based on quantum Fourier transform \cite{60}, the generalized CNOT operator can be implemented in quantum circuits.

\begin{figure}[htbp]
  \centering
  \includegraphics[width=0.48\textwidth]{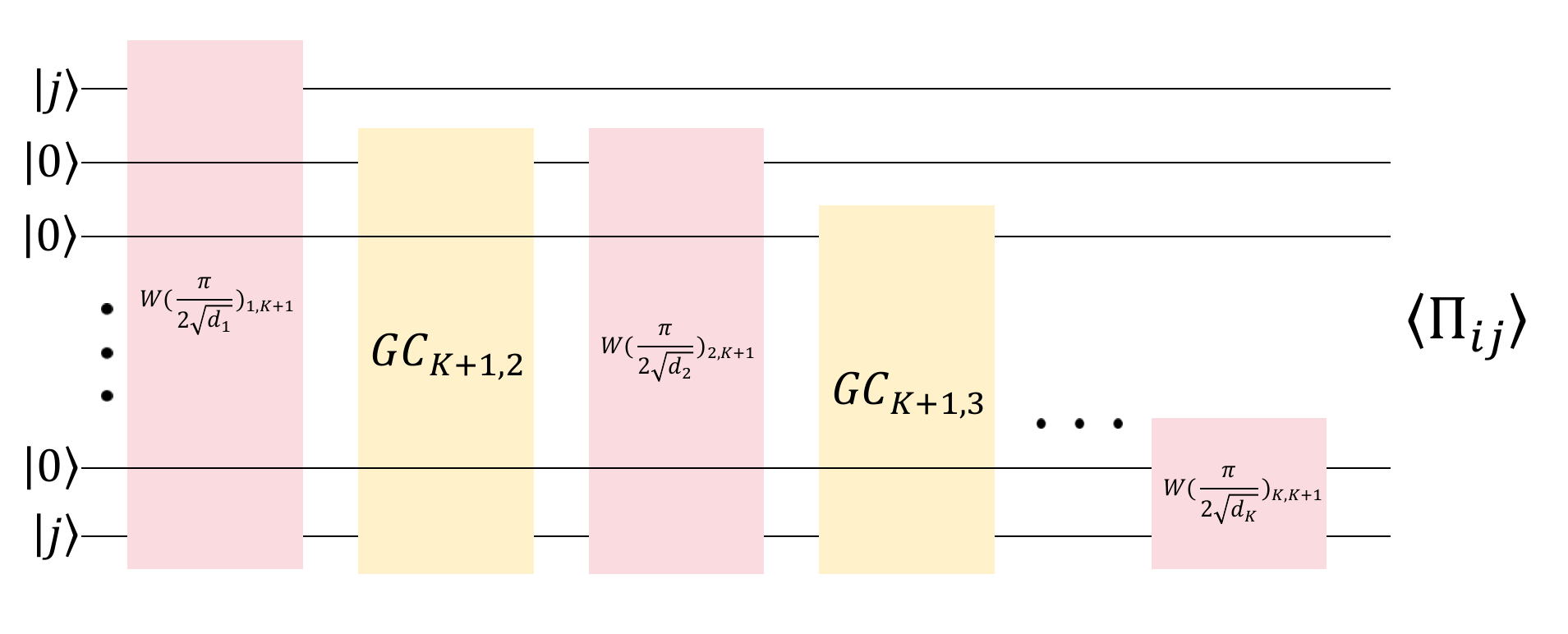}\\
  \caption{The quantum circuit diagram for the matrix multiplication algorithm. The initial state is $|j0\cdots0j\rangle_{1,2,\cdots,K,K+1}$. Each pink rectangle represents a hybrid quantum walk operator, and the yellow rectangle signifies a generalized CNOT operator. The subscript indicates the position where the operator acts. The output is the observed value of the projection operator $\Pi_{ij}$.}
  \label{fig:26}
\end{figure}

To calculate $C_{ij}=(A^{(K)}\cdots A^{(1)})_{ij}$, we design a quantum process as shown in FIG.~\ref{fig:26}. 
The initial quantum state is 
\begin{align}
|\Psi_0\rangle=|j0\cdots0j\rangle_{1,2,\cdots,K,K+1}.  
\end{align}
After the first hybrid quantum walk and the first generalized CNOT operator,
\begin{align}
|\Psi_1\rangle&= GC_{K+1,2}W^{(1)}(\frac{\pi}{2\sqrt{d_1}})_{1,K+1}|\Psi_0\rangle\nonumber\\
&=|j\rangle\sum_{p_1=0}^{n-1}\frac{A_{p_1j}^{(1)}}{\sqrt{d_1}}|p_10\cdots0p_1\rangle,
\end{align}
where the hybrid quantum walk $W^{(1)}(\frac{\pi}{2\sqrt{d_1}})_{1,K+1}$ acts on particles $1$ and $K+1$.
After the $l$-th hybrid quantum walk and generalized CNOT operator, the quantum state is 
\begin{align}
|\Psi_l\rangle&= GC_{K+1,l+1}W^{(l)}(\frac{\pi}{2\sqrt{d_l}})_{l,K+1}|\Psi_{l-1}\rangle\nonumber\\
&=|j\rangle\sum_{p_1,\cdots, p_l=0}^{n-1}\frac{A_{p_lp_{l-1}}^{(l)}\cdots A_{p_1j}^{(1)}}{\sqrt{d_l\cdots d_1}}|p_1\cdots p_l0\cdots0p_l\rangle.
\end{align}
Then the final quantum state $|\Psi_f\rangle$ is
\begin{align}
|j\rangle\sum_{p_1,\cdots, p_K=0}^{n-1}\frac{A_{p_Kp_{K-1}}^{(K)}\cdots A_{p_1j}^{(1)}}{\sqrt{d_K\cdots d_1}}|p_1\cdots p_K\rangle.
\end{align}
Given a project operator $\Pi_{ij}=|j\rangle\langle j|\otimes I\otimes\cdots\otimes I\otimes|i\rangle\langle i|$, and because 
\begin{align}
(A_{p_Kp_{K-1}}^{(K)}\cdots A_{p_1j}^{(1)})^2=A_{p_Kp_{K-1}}^{(K)}\cdots A_{p_1j}^{(1)},
\end{align}
then 
\begin{align}
C_{ij}=d_K\cdots d_1|\Pi_{ij}|\Psi_f\rangle|^2.
\end{align}
The specific calculation process can be found in Appendix B. The specific algorithm process is described in \textbf{Algorithm 2}.
\begin{algorithm}[H]
\caption{Quantum Matrix Product Computation}
\label{alg:matrix_product}
\begin{algorithmic}[1]
\Require Regular graphs $G_l=(V_l,E_l)$ with adjacency matrices $A^{(l)}$, $l=1,\dots,K$
\Ensure Matrix product entries $C_{ij}=(A^{(K)}\cdots A^{(1)})_{ij}$

\State \textbf{Initialization:}
\State Prepare initial state $|\Psi_0\rangle=|j0\cdots0j\rangle_{1,2,\cdots,K,K+1}$

\State \textbf{Quantum Walk Processing:}
\For{$l=1$ \textbf{to} $K$}
    \State Apply hybrid quantum walk operator:
    \[
    W^{(l)}(\tfrac{\pi}{2\sqrt{d_l}})_{l,K+1} = \sum_{k=0}^{n-1}|k\rangle\langle k|\otimes e^{-iS_k^{(l)}\tfrac{\pi}{2\sqrt{d_l}}}
    \]
    \State Apply generalized CNOT gate:
    \[
    GC_{K+1,l+1} = \sum_{i,j=0}^{n-1}|i\rangle\langle i|\otimes|(j+i)\ \mathrm{mod}\ n\rangle\langle j|
    \]
    \State Update quantum state:
    \[
    |\Psi_l\rangle = GC_{K+1,l+1}W^{(l)}(\tfrac{\pi}{2\sqrt{d_l}})_{l,K+1}|\Psi_{l-1}\rangle
    \]
\EndFor

\State \textbf{Measurement:}
\State Construct projection operator:
\[
\Pi_{ij} = |j\rangle\langle j|\otimes I^{\otimes K-1}\otimes|i\rangle\langle i|
\]
\State Measure amplitude:
\[
\hat{a}_{ij} \leftarrow \text{AmplitudeEstimation}(\Pi_{ij}|\Psi_f\rangle)
\]
\State Calculate matrix entry:
\[
\hat{C}_{ij} = d_K\cdots d_1 \cdot |\hat{a}_{ij}|^2
\]
\end{algorithmic}
\end{algorithm}
Because $C_{ij}$ is a integer, the estimate $\hat{C}_{ij}$ should satisfy 
\begin{align}
|\hat{C}_{ij}-C_{ij}|<\frac{1}{2}. 
\end{align}
And the amplitude estimate $\hat{a}_{ij}$ of $|\Pi_{ij}|\Psi_f\rangle|$ should satisfy $|\hat{a}_{ij}|<1$, then 
\begin{align}
&|\hat{a}_{ij}^2-|\Pi_{ij}|\Psi_f\rangle|^2|\nonumber\\
&=|\hat{a}_{ij}+|\Pi_{ij}|\Psi_f\rangle|||\hat{a}_{ij}-|\Pi_{ij}|\Psi_f\rangle||\nonumber\\
&\leq2|\hat{a}_{ij}-|\Pi_{ij}|\Psi_f\rangle||.
\end{align}
Therefore, when $|\hat{a}_{ij}-|\Pi_{ij}|\Psi_f\rangle||<\frac{1}{4d_K\cdots d_1}$, $|\hat{C}_{ij}-C_{ij}|<\frac{1}{2}$. According to the amplitude estimation algorithm \cite{53}, the complexity of estimating $|\Pi_{ij}|\Psi_f\rangle|$ should be $O(d_K\cdots d_1)$. Then the complexity of obtaining $C_{ij}$ is $O(d_K\cdots d_1)$. So the complexity of calculating $A^{(K)}\cdots A^{(1)}$ is $O(n^2d_K\cdots d_1)$. We leverage the structure of the adjacency matrix of regular graphs to ensure that the time complexity solely depends on the number and the degree of the vertices, without depending on the precision $\varepsilon$. Therefore, when the degree of the vertices is a number unrelated to $n$ in every graph, the complexity of our algorithm is $O(n^2)$. However, for the algorithms in \cite{54} and \cite{56}, when $\delta=\frac{1}{2}$ and matrices $A$ and $B$ are adjacency matrices of regular graphs with degree $d_1$ and $d_2$, $\varepsilon^{-1}>2n^2d_1d_2$. Thus, the complexity of the algorithms in \cite{54} and \cite{56} are $O(n^4d_1d_2)$ and $O(n^6d_1^2d_2^2)$, respectively.
Both our algorithm and the algorithm in \cite{56} encode the product results into the amplitudes of quantum states. However, it is not easy for the algorithm in \cite{56} to encode the product of multiple matrices into the amplitudes of quantum states, whereas our algorithm can achieve this. Therefore, we also provide a new encoding method for the product of a certain class of matrices. Compared with the fastest classical matrix multiplication algorithm, when $d_K\cdots d_1<n^{0.371552}$, our algorithm can demonstrate its advantages. Although the classical algorithm for calculating the product of two sparse matrices has lower complexity by using the sparse structure of the matrix, the matrix will gradually become non-sparse in the process of calculating the product of several sparse matrices, and the sparse matrix product algorithm will gradually fail. Therefore, the classic sparse matrix product algorithm is sometimes not applicable to calculating the product of several sparse matrices. However, our algorithm can calculate the product of several sparse matrices and is not affected by the number of matrices. Since we can obtain $C_{ij}$ with the complexity $O(d_K\cdots d_1)$, we can calculate $tr(A^{(K)}\cdots A^{(1)}), K>2$, with the complexity $O(nd_K\cdots d_1)$ instead of $O(n^{2.371552})$. Thus when we want to obtain the trace of $A^{(K)}\cdots A^{(1)}$ and $d_K\cdots d_1<n^{1.371552}$, our algorithm can demonstrate advantages.

As a key application, our matrix multiplication algorithm enables efficient triangle counting in networks—a fundamental task for analyzing network properties such as transitivity and clustering coefficients \cite{44,51}. The standard algebraic approach computes $\frac{\text{tr}(A^3)}{6}$ through the adjacency matrix $A$ \cite[Corollary 8.1.3]{61}. Alternative enumeration-based methods include the $O(m^{3/2})$-time algorithm by Itai and Rodeh \cite{50}, where $m$ is the number of edges, which incurs high overhead from graph data structure modifications, limiting its practical utility. Other methods like NodeIterator ($O(n \cdot d_{\text{max}}^2)$) and EdgeIterator ($O(m \cdot d_{\text{max}})$) achieve comparable $O(m^{3/2})$ complexity bounds \cite{51}. Recent studies on regular graphs \cite{52} focus primarily on theoretical lower bounds for triangle counts rather than providing exact counting algorithms, further highlighting the need for efficient algebraic solutions.

Thus, we apply our quantum algorithm for the triangle counting in $d$-regular graphs $G=(V, E)$, with the adjacency matrix $A$. Given a vertex $k$ of the graph $G$, utilize the above matrix multiplication algorithm to obtain the $k$-th diagonal element of $A^3$, thereby yielding the number of triangles containing vertex $k$ through Lemma 8.1.2 of \cite{61}. We can compute the number of triangles in the graph by iterating over all vertices. The entire process requires $n$ invocations of the above procession, thus the time complexity is $O(nd^3)$. Because the time complexity of the triangle counting algorithm based on matrix multiplication is $O(n^{2.371552})$, when $d<n^{0.457184}$, our algorithm can be better than the algorithm based on matrix multiplication. Although our algorithm has a complexity that is higher by a factor of d compared to the complexity of enumeration-based algorithms, the dependency on the number of vertices remains the same. Furthermore, our use of quantum processes instead of enumeration makes it more suitable for large-scale regular networks. 

\begin{figure*}[htpb]  
\centering  
\begin{subfigure}[htbp]{0.25\textwidth}  
\centering          \includegraphics[width=\textwidth]{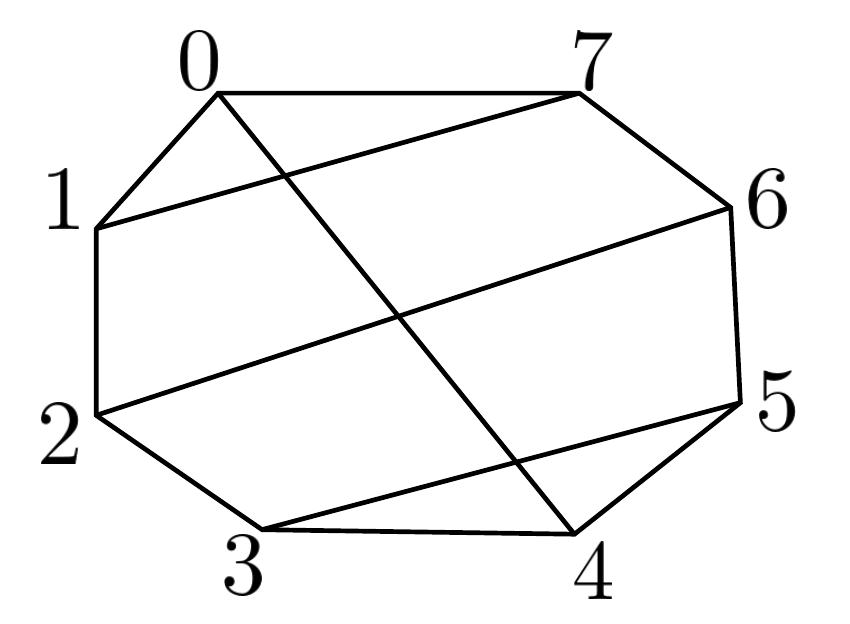} 
\caption{}  
\label{fig:27}  
\end{subfigure} 
\begin{subfigure}[htbp]{0.7\textwidth}  
\centering        \includegraphics[width=\textwidth]{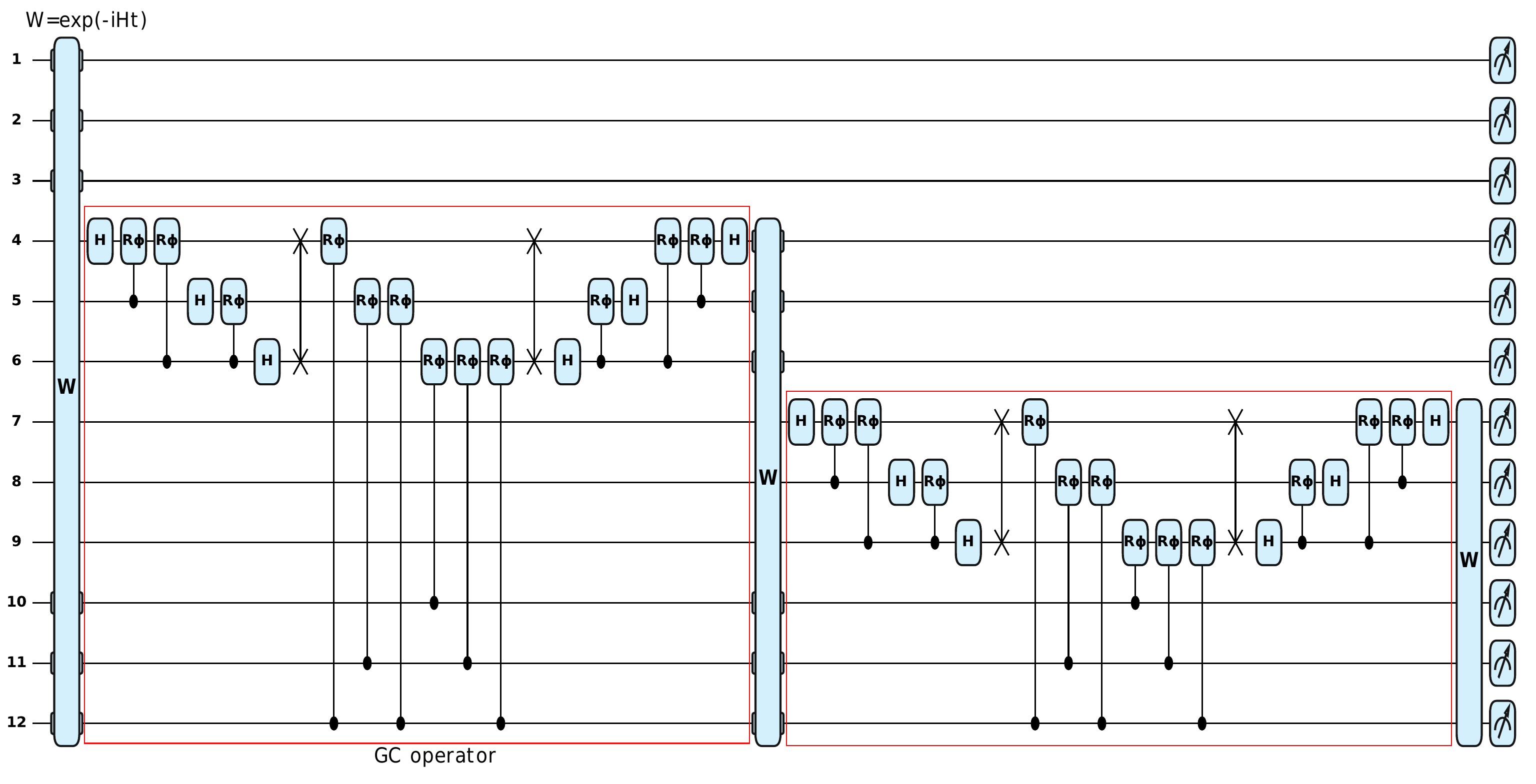} 
\caption{}  
\label{fig:28}  
\end{subfigure} 
\caption{(a) An 8-vertex 3-regular graph $G$. (b) The quantum circuit to calculate $A^3_{00}=$ for $G$. The blue rectangle $W$ represents the
hybrid quantum walk on the graph $G$ which can be implemented by $W=\exp{(iH\frac{\pi}{2\sqrt{3}})}$ in PennyLane. The GC operator can be implemented by the quantum adder based on quantum Fourier transform \cite{60}.}
\label{fig:2728}  
\end{figure*}

To verify the correctness and feasibility of our matrix multiplication algorithm for solving the triangle counting problem, we designed an algorithmic circuit using PennyLane with an example of an 8-vertex 3-regular graph as shown in FIG.~\ref{fig:27} and obtained experimental results. For example, through the experiment circuit as shown in FIG.~\ref{fig:28}, we can obtain the experimental result of $\langle\Pi_{00}\rangle\approx0.0741$, thus $A^3_{00}=0.0741\times3^3=2.0007\approx2$. Thus the number of the triangles contain vertex $0$ is $\frac{A^3_{00}}{2}=1$.
We can choose the initial state as $|i0\cdots0i\rangle$ to calculate $A^3_{ii}$, then we can get $\frac{tr(A^3)}{6}$.

\section{\MakeUppercase\expandafter{\romannumeral7}. conclusions}\label{s7}

In this paper, we propose a hybrid quantum walk model. In this new model, we utilize the continuous quantum walk model to construct the conditional shift operator in coin-based quantum walks, such that each step of the quantum walk is a continuous quantum walk on a subgraph controlled by a coin. Our hybrid model retains the coin-controlled functionality of discrete quantum walks and introduces the evolution mechanism of continuous quantum walks on graphs. Furthermore, the new model is more general, allowing existing walk models to be special cases. We also characterize the dynamic evolution, probability distribution, standard deviation, and entanglement entropy of this hybrid model on the 2-vertex circles, star graphs, and lines. Leveraging the new model, we achieve perfect state transfer on general connected graphs. We also present a quantum matrix multiplication algorithm for adjacency matrices of regular graphs. The time complexity of the new algorithm is $O(n^2d_1\cdots d_K)$ instead of $O(n^{2.371552})$ or depending on the precision $\varepsilon$. And we apply the algorithm to the triangle counting problem and conducted experimental verification on an 8-vertex 3-regular graph using the quantum simulation on PennyLane. Quantum walk models have already played significant roles in numerous fields of quantum information technology, so our hybrid walk model is also expected to offer greater advantages in these areas. We will continue to explore the applications of this hybrid model in other domains as well.

\section{ACKNOWLEDGMENTS}

\begin{acknowledgments}
The research is supported by the National Key R\&D Program of China under Grant No. 2023YFA1009403, the National Natural Science Foundation special project of China (Grant No.12341103), the National Natural Science Foundation of China (Grant No.62372444).
\end{acknowledgments}

\section{APPENDIX A: Factors affecting the hybrid quantum walk on lines.}

For hybrid quantum walks, similar to classical discrete and continuous quantum walks, some factors affect the evolution of the walk, such as the choice of the initial coin state, the number of steps, or the duration of the walk. In Appendix A, we primarily focus on the three-label hybrid quantum walk on a line and investigate how the evolution time for a single-step hybrid quantum walk, as well as the phase and distribution of the initial coin state, affect the evolution of this quantum walk. The probability distribution at the nodes on the line will exhibit the results of these influencing factors.

\subsection{Evolutionary time of a single-step hybrid quantum walk}

In this section, we primarily investigate how the evolution time of a single-step hybrid quantum walk and the choice of the initial state influence the probability distribution of the final position state across the nodes. 

\begin{figure}[htpb]  
    \centering  
    \begin{subfigure}[htbp]{0.46\textwidth}  
        \centering  \includegraphics[width=\textwidth]{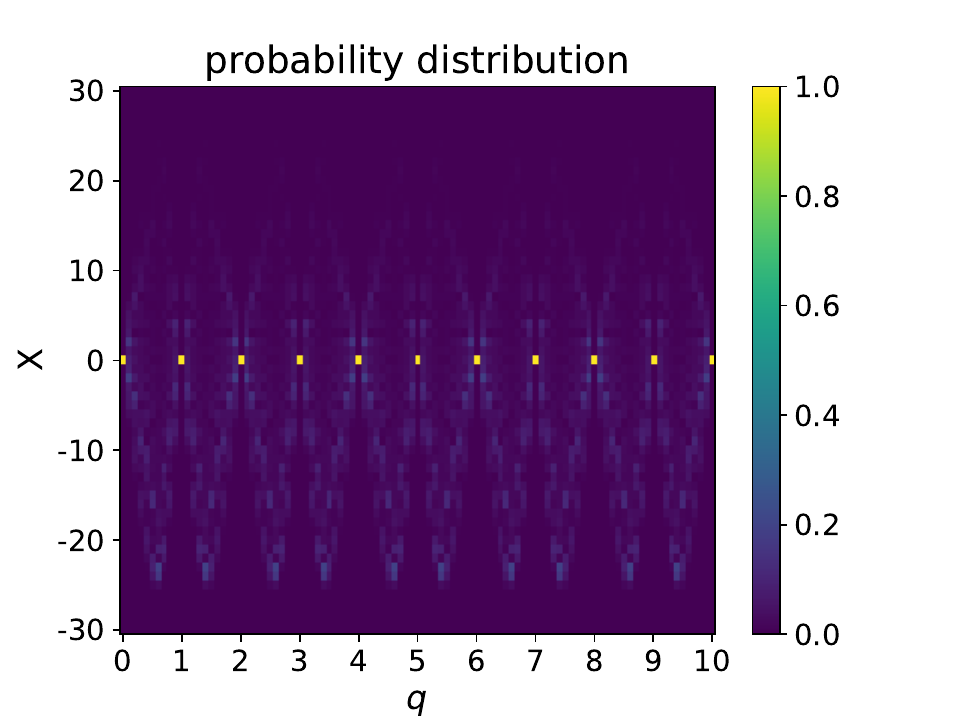}  
        \subcaption{}  
        \label{fig:17}  
    \end{subfigure}   
    
    \begin{subfigure}[htbp]{0.46\textwidth}  
        \centering          \includegraphics[width=\textwidth]{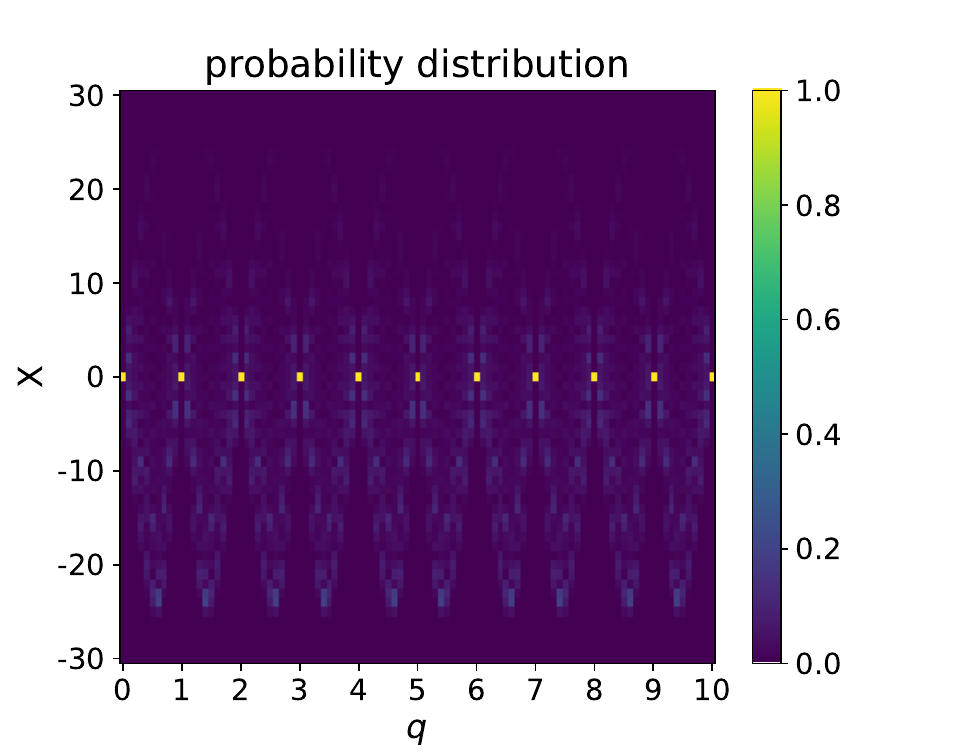}  
        \subcaption{}  
        \label{fig:18}  
    \end{subfigure}  
    
    \begin{subfigure}[htbp]{0.46\textwidth}  
        \centering  \includegraphics[width=\textwidth]{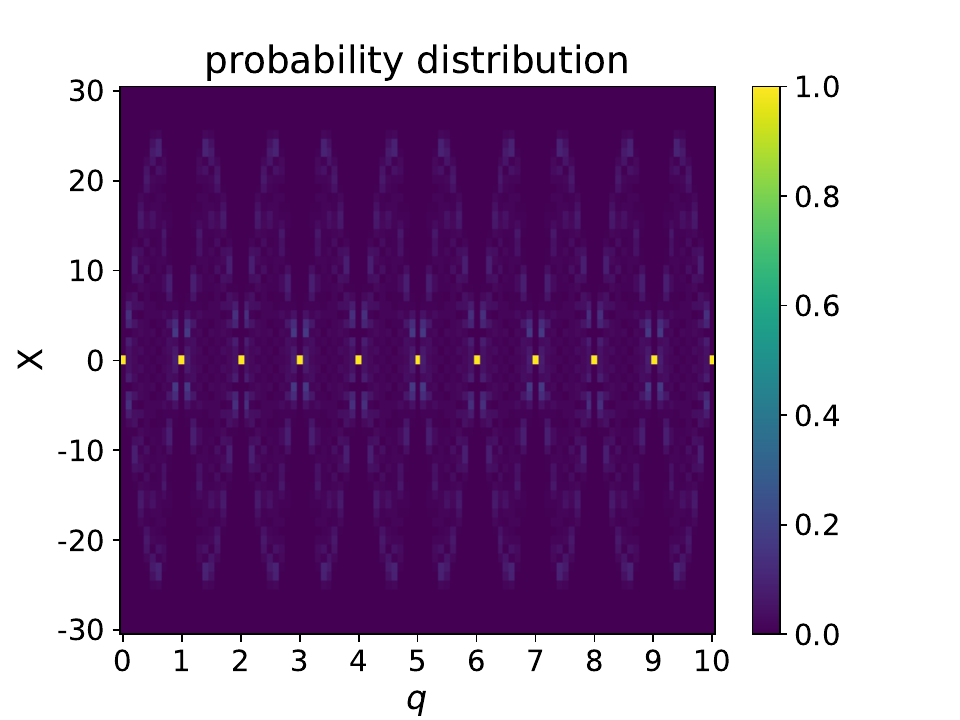}  
        \subcaption{}  
        \label{fig:19}  
    \end{subfigure}   
    
    \caption{Probability distribution of the position state on line nodes $X$, starting from the initial state (a) $|0\rangle\otimes|0\rangle$, (b) $\frac{|0\rangle+|1\rangle}{\sqrt{2}}\otimes|0\rangle$, and (c) $\frac{|0\rangle+|1\rangle+|2\rangle}{\sqrt{3}}\otimes|0\rangle$, varying with evolutionary time $q\pi$ of a single-step hybrid quantum walk.}     
    \label{fig:fig A1}  
\end{figure}

The hybrid quantum walk operator is 
\begin{align}
   &W(q)=e^{-iHq\pi}\otimes(C\otimes I)\nonumber\\
&=(|0\rangle\langle0|\otimes\sum_{m=3k,k\in\mathbb{Z}}R_x(2q\pi)_{m,m+1}\nonumber\\&+|1\rangle\langle1|\otimes\sum_{m=3k+1,k\in\mathbb{Z}}R_x(2q\pi)_{m,m+1}\nonumber\\&+|2\rangle\langle2|\otimes\sum_{m=3k+2,k\in\mathbb{Z}}R_x(2q\pi)_{m,m+1})\otimes(C\otimes I), 
\end{align}
 where $H$ and $C$ are defined in the main text. And 30 step quantum walks were carried out. Key findings are summarized below:

\textbf{Regulation effect of parameter $q$}: 1) Localization Phenomena: When $ q\in\mathbb{Z}^+$, positional states remain strictly localized at the initial node (Fig.~\ref{fig:fig A1}). Here, the evolution operator reduces to a global identity operation, preventing inter-node state diffusion. 2) Maximum Diffusion Condition: At \( q = k\pi + \frac{\pi}{2} \, (k \in \mathbb{N}) \), positional states achieve maximal diffusion to the farthest nodes. This arises from the rotation operator: $R_x(2q\pi)_{m,m+1} = -iT_{m,m+1}$, enabling complete state exchange between adjacent nodes (see Fig.~\ref{fig:fig A1}).

\textbf{Directional Control via Initial Coin State Symmetry}: 1) \textit{Single-Basis State}: For $|0\rangle \otimes |0\rangle$, positional states exhibit left-biased diffusion; 2) \textit{Two-Basis Superposition}: With $\frac{|0\rangle + |1\rangle}{\sqrt{2}} \otimes |0\rangle$, right-side node accessibility increases; 3) \textit{Three-Basis Symmetric State}: The state $\frac{|0\rangle + |1\rangle + |2\rangle}{\sqrt{3}} \otimes |0\rangle$ induces symmetric bilateral diffusion. Dynamic analysis reveals that the basis states in the initial coin state correspond to independent diffusion channels on the three-label line. Multi-component superposition generates quantum interference effects that modulate diffusion symmetry, with increased basis-state diversity enhancing bilateral symmetry.

\subsection{The distribution and phase of the initial coin state}

We next investigate the impact of the proportion and the phase of different basis states in the initial coin state on the probability distribution of the final positional state.

\begin{figure}[htpb]  
    \centering  
    \begin{subfigure}[htbp]{0.46\textwidth}  
        \centering  \includegraphics[width=\textwidth]{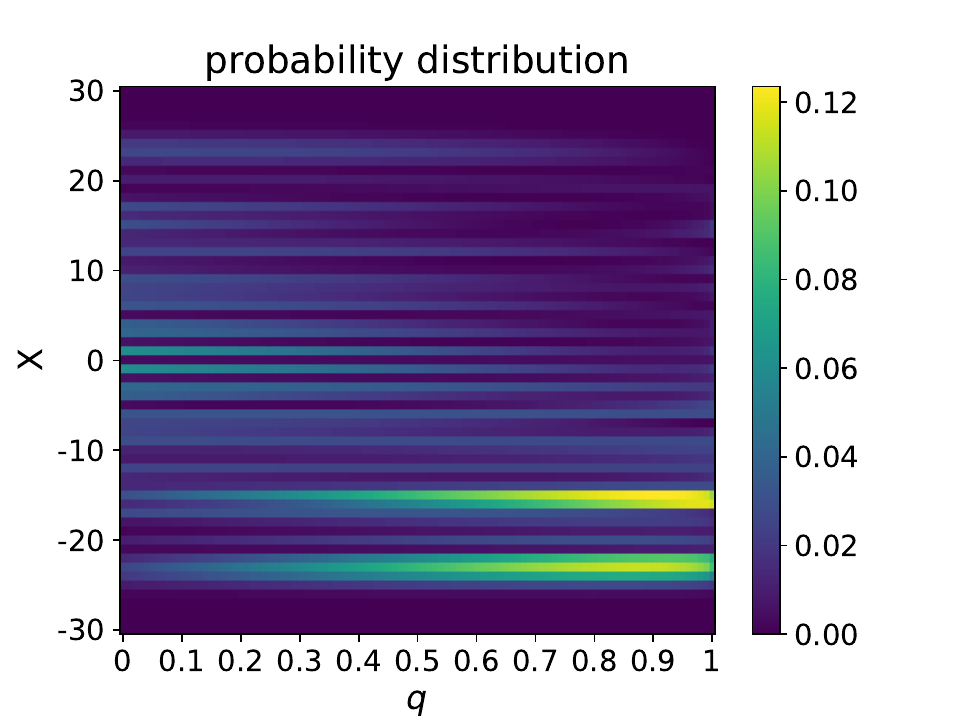}  
        \caption{}  
        \label{fig:22}  
    \end{subfigure}  
    
    \begin{subfigure}[htbp]{0.46\textwidth}  
        \centering          \includegraphics[width=\textwidth]{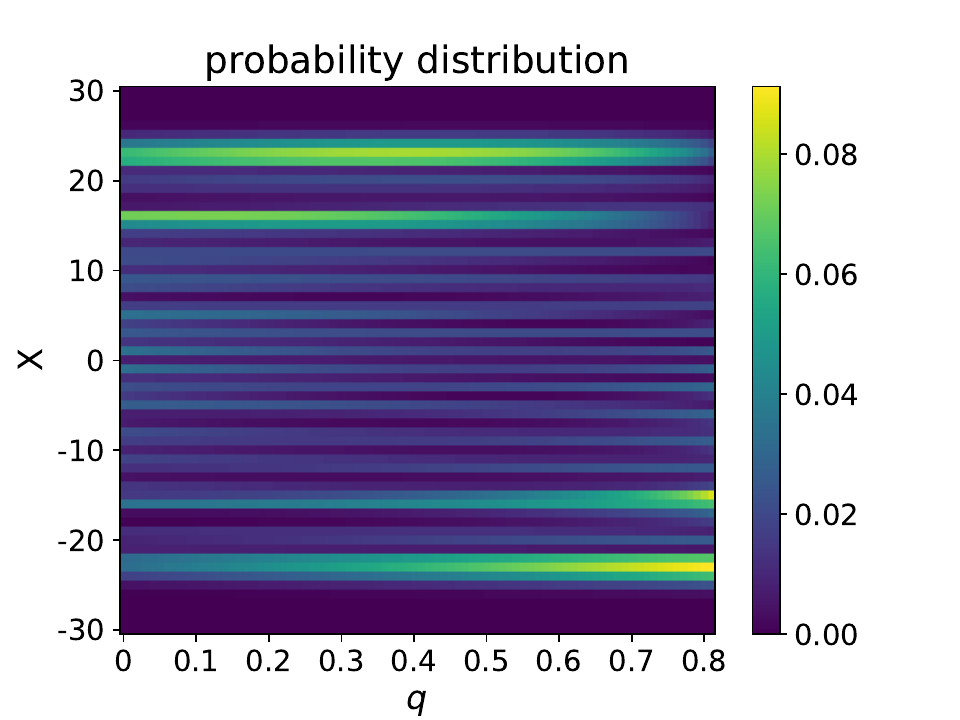} 
        \caption{}  
        \label{fig:23}  
    \end{subfigure}  

    \caption{Probability distribution of the position state on line nodes $X$, starting from the initial state (a) $(q|0\rangle+\sqrt{1-q^2}|1\rangle)\otimes|0\rangle$, (b) $(\frac{|0\rangle+\sqrt{3}q|1\rangle+\sqrt{2-3q^2}|2\rangle}{\sqrt{3}})\otimes|0\rangle$.}  
    \label{fig:fig A2}  
\end{figure}

\begin{figure}[htpb]  
    \centering  
    \begin{subfigure}[htbp]{0.46\textwidth}  
        \centering  \includegraphics[width=\textwidth]{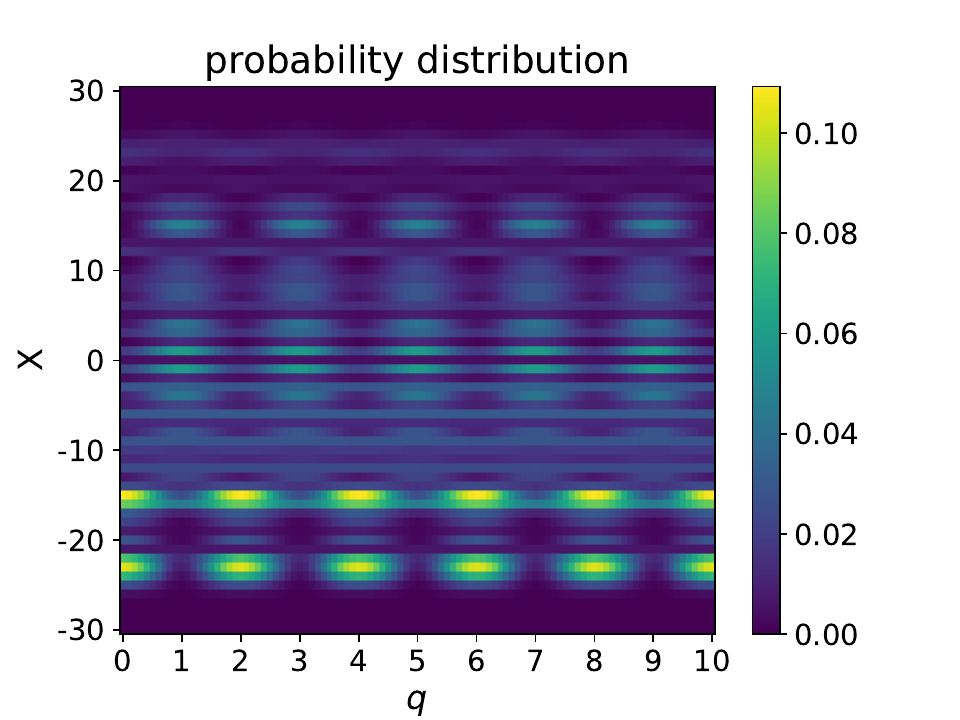}  
        \caption{}  
        \label{fig:20}  
    \end{subfigure}  
    
    \begin{subfigure}[htbp]{0.46\textwidth}  
        \centering          \includegraphics[width=\textwidth]{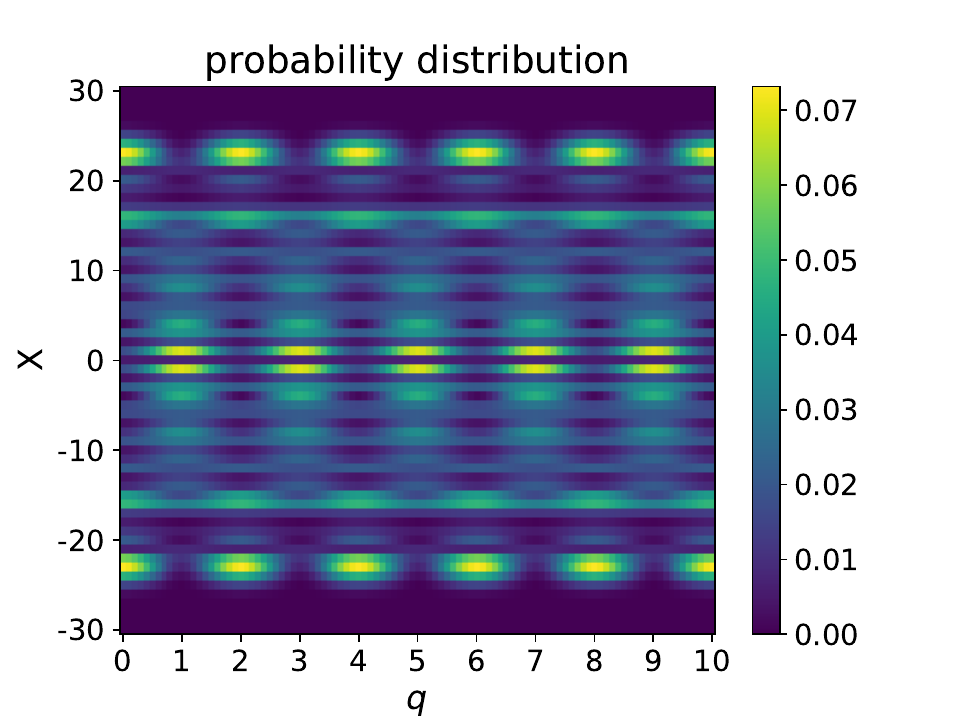} 
        \caption{}  
        \label{fig:21}  
    \end{subfigure}  

    \caption{Probability distribution of the position state on line nodes $X$, starting from the initial state (a) $\frac{|0\rangle+e^{-iq\pi}|1\rangle}{\sqrt{2}}\otimes|0\rangle$, (b) $\frac{|0\rangle+e^{-iq\pi}|1\rangle+|2\rangle}{\sqrt{3}}\otimes|0\rangle$.}  
    \label{fig:fig A3}  
\end{figure}

The hybrid quantum walk operator is defined as $W=e^{-i\frac{3\pi}{2}H}\otimes(C\otimes I)$, with initial states $(q|0\rangle+\sqrt{1-q^2}|1\rangle)\otimes|0\rangle$ and $\frac{|0\rangle+\sqrt{3}q|1\rangle+\sqrt{2-3q^2}|2\rangle}{\sqrt{3}}\otimes|0\rangle$. Numerical results for 30-step walks are shown in FIG.~\ref{fig:fig A2}. Key findings are summarized below: \textbf{Asymmetry Modulation}: For initial state $(q|0\rangle+\sqrt{1-q^2}|1\rangle)\otimes|0\rangle$, $q$ varies from $0$ to $1$, resulting in a gradual change in the final positional state from a symmetrical spread to a left-biased localization as can be seen from  FIG.~\ref{fig:22}.
\textbf{Lateral reversal effect}: FIG.~\ref{fig:23} demonstrates that as $q$ varies from 0 to $\sqrt{\frac{2}{3}}$, the initial coin state $\frac{|0\rangle+\sqrt{3}q|1\rangle+\sqrt{2-3q^2}|2\rangle}{\sqrt{3}}$ changes from $\frac{|0\rangle+\sqrt{2}|2\rangle}{\sqrt{3}}$ to $\frac{|0\rangle+\sqrt{2}|1\rangle}{\sqrt{3}}$, resulting in a gradual change in the final positional state from a distribution predominantly on the right side of the initial node to the left side. \textbf{Phase-controlled diffusion}: FIG.~\ref{fig:fig A3} demonstrates the influence of local phase modulation on positional state dynamics using initial coin states: $\frac{|0\rangle + e^{-iq\pi}|1\rangle}{\sqrt{2}} \otimes |0\rangle$ and $\frac{|0\rangle + e^{-iq\pi}|1\rangle + |2\rangle}{\sqrt{3}} \otimes |0\rangle$. As can also be seen from FIG.~\ref{fig:fig A3}, the more basic components in the initial coin state, the more symmetrical the diffusion of the position state towards both sides. However,  something more peculiar is that when $q$ is even, the probability value of the position state is maximized at farther reachable nodes, but when $q$ is odd, the probability of the position state achieves its maximum value near the initial node as shown in FIG.~\ref{fig:21}. This indicates that when $q$ is even or odd, it increases or decreases the diffusion of the position state. Corresponding to the initial coin state, when $q$ is even or odd, the coefficient of component $|1\rangle$ is $\pm1$, so $\pm1$ respectively increases or decreases the diffusion of the position state.

\section{APPENDIX B: The specific calculation process of adjacency matrices of regular graphs multiplication algorithms.}

The initial quantum state is 
\begin{align}
|\Psi_0\rangle=|j0\cdots0j\rangle_{1,2,\cdots,K,K+1}.  
\end{align} 
Thus 
\begin{align}
&W^{(1)}(t)_{1,K+1}|\Psi_0\rangle\\
&=\sum_{k=0}^{n-1}|k\rangle_1\langle k|\otimes I\otimes\cdots\otimes I\otimes e^{-iS_k^{(1)}t}|j0\cdots0j\rangle_{1,2,\cdots,K,K+1}\\
&=|j0\cdots0\rangle e^{-iS_j^{(1)}t}|j\rangle\\
&=|j0\cdots0\rangle(\frac{e^{-i\sqrt{d_1}t}}{\sqrt{2}}|\lambda_{j,+}^{(1)}\rangle-\frac{e^{i\sqrt{d_1}t}}{\sqrt{2}}|\lambda_{j,-}^{(1)}\rangle\\
&+|j\rangle-\frac{|\lambda_{j,+}^{(1)}\rangle-|\lambda_{j,-}^{(1)}\rangle}{\sqrt{2}})\\
&=|j0\cdots0\rangle(\frac{e^{-i\sqrt{d_1}t}}{\sqrt{2}}|\lambda_{j,+}^{(1)}\rangle-\frac{e^{i\sqrt{d_1}t}}{\sqrt{2}}|\lambda_{j,-}^{(1)}\rangle).
\end{align}
When $t=\frac{\pi}{2\sqrt{d_1}}$, we can get 
\begin{align}
\frac{e^{-i\sqrt{d_1}t}}{\sqrt{2}}|\lambda_{j,+}^{(1)}\rangle-\frac{e^{i\sqrt{d_1}t}}{\sqrt{2}}|\lambda_{j,-}^{(1)}\rangle=\sum_{p_1=0}^{n-1}\frac{A_{p_1j}^{(1)}}{\sqrt{d_1}}|p_1\rangle.
\end{align}
Thus 
\begin{align}
W^{(1)}(\frac{\pi}{2\sqrt{d_1}})_{1,K+1}|\Psi_0\rangle=|j\rangle\sum_{p_1=0}^{n-1}\frac{A_{p_1j}^{(1)}}{\sqrt{d_1}}|00\cdots0p_1\rangle.
\end{align}
To perform the second step of quantum walk $W^{(2)}(\frac{\pi}{2\sqrt{d_2}})_{2,K+1}$, we need to use a generalized CNOT operator $GC_{K+1,2}$ to act on particles 2 and K+1 to obtain $|\Psi_1\rangle=|j\rangle\sum_{p_1=0}^{n-1}\frac{A_{p_1j}^{(1)}}{\sqrt{d_1}}|p_10\cdots0p_1\rangle$. 
Thus 
\begin{align}
|\Psi_2\rangle&= GC_{K+1,3}W^{(2)}(\frac{\pi}{2\sqrt{d_2}})_{2,K+1}|\Psi_{1}\rangle\nonumber\\
&=|j\rangle\sum_{p_1, p_2=0}^{n-1}\frac{A_{p_2p_1}^{(2)}A_{p_1j}^{(1)}}{\sqrt{d_2d_1}}|p_1p_20\cdots0p_2\rangle.
\end{align}
And after the $l$-th hybrid quantum walk and generalized CNOT operator, the quantum state is 
\begin{align}
|\Psi_l\rangle&= GC_{K+1,l+1}W^{(l)}(\frac{\pi}{2\sqrt{d_l}})_{l,K+1}|\Psi_{l-1}\rangle\nonumber\\
&=|j\rangle\sum_{p_1,\cdots, p_l=0}^{n-1}\frac{A_{p_lp_{l-1}}^{(l)}\cdots A_{p_1j}^{(1)}}{\sqrt{d_l\cdots d_1}}|p_1\cdots p_l0\cdots0p_l\rangle.
\end{align}
Then the final quantum state $|\Psi_f\rangle$ is
\begin{align}
|j\rangle\sum_{p_1,\cdots, p_K=0}^{n-1}\frac{A_{p_Kp_{K-1}}^{(K)}\cdots A_{p_1j}^{(1)}}{\sqrt{d_K\cdots d_1}}|p_1\cdots p_K\rangle.
\end{align}
Thus $|\Pi_{ij}|\Psi_f\rangle|^2=\sum_{p_1,\cdots, p_{K-1}=0}^{n-1}\frac{(A_{ip_{K-1}}^{(K)}\cdots A_{p_1j}^{(1)})^2}{d_K\cdots d_1}$. since $(A_{p_Kp_{K-1}}^{(K)}\cdots A_{p_1j}^{(1)})^2=A_{p_Kp_{K-1}}^{(K)}\cdots A_{p_1j}^{(1)}$, we can get $|\Pi_{ij}|\Psi_f\rangle|^2=\sum_{p_1,\cdots, p_{K-1}=0}^{n-1}\frac{A_{ip_{K-1}}^{(K)}\cdots A_{p_1j}^{(1)}}{d_K\cdots d_1}$. Thus $C_{ij}=(A^{(K)}\cdots A^{(1)})_{ij}=d_K\cdots d_1|\Pi_{ij}|\Psi_f\rangle|^2$.

\bibliography{ref}

\end{document}